\documentclass[a4paper,11pt]{article}
\usepackage{jinstpub} % for details on the use of the package, please see the JINST-author-manual
\usepackage{lineno}
\usepackage{siunitx}
\usepackage{multirow}
\usepackage{subcaption}
\usepackage{booktabs} 
\usepackage{array} 

\title{\boldmath Surface Commissioning and Performance of the sMDT Muon Chambers for the ATLAS HL-LHC Upgrade}

\author[a,1]{Jiajin Ge,\note{Corresponding authors.}}
\author[b,1]{Elena Voevodina,}
\author[b,1]{Hubert Kroha,}
\author[a, c]{Tatiana Azaryan,}
\author[d]{Kwok Ching Cheung,}
\author[a]{Tiesheng Dai,}
\author[a]{Edward Diehl,}
\author[a]{Claudio Ferretti,}
\author[a]{Yuxiang Guo,}
\author[b]{Oliver Kortner,}
\author[a]{Chihao Li,}
\author[d]{Chun Kit Lo,}
\author[b]{Nick Meier,}
\author[a]{Emmett Salzer,}
\author[a]{Can Suslu,}
\author[e]{Cecilia Vanesa Imthurn,}
\author[c]{Samuel Hugo Venetianer,}
\author[a]{Curtis Weaverdyck,}
\author[a, f]{Chuanshun Wei,}
\author[b]{Bastian Michael Wesely,}
\author[a]{Ruslan Yakubovych,}
\author[a]{Bing Zhou,}
\author[a]{Junjie Zhu,}
\author[b]{J\"org Zimmermann}

\affiliation[a]{Department of Physics, University of Michigan, Ann Arbor, MI, United States of America}
\affiliation[b]{Max Planck Institute for Physics, Boltzmannstraße 8,  Garching, Germany}
\affiliation[c]{Department of Physics and Astronomy, Tufts University, Medford, MA, United States of America}
\affiliation[d]{Department of Physics, Chinese University of Hong Kong, Shatin, N.T., Hong Kong}
\affiliation[e]{Department of Physics and Astronomy, Michigan State University, East Lansing, MI, United States of America}
\affiliation[f]{Department of Modern Physics, University of Science and Technology of China, Hefei 230026, China}

\emailAdd{jiajin.ge@cern.ch, elena.voevodina@cern.ch, kroha@mpp.mpg.de}

\abstract{
To improve the first-level muon trigger efficiency at the HL-LHC, the Monitored Drift Tube (MDT) chambers in the inner small sectors of the ATLAS Barrel Muon Spectrometer will be replaced with integrated modules of small-Diameter Muon Drift Tube (sMDT) and Resistive Plate Chambers (RPCs).
This paper reports on the surface commissioning of 102 new sMDT chambers at CERN in 2025 following the installation of the final front-end electronics, including the new Amplifier-Shaper-Discriminator (ASD), Time-to-Digital Converter (TDC) chips, and Chamber Service Module (CSM) developed for the HL-LHC. The commissioning included measurements of gas leak rates and high-voltage dark currents, tests of the integrated planarity monitoring system, noise characterization, and measurements of detector efficiency and spatial resolution using cosmic rays. Fewer than 0.1\% of the 49152 tubes were found to be non-functional due to broken sense wires or gas leaks. The chambers exceed the ATLAS design requirements, with gas leak rates a factor of five below the specified limit of
$9.3\times 10^{-3}~\mathrm{mbar\cdot liter/s}$ per chamber, dark currents of only around 0.2~nA per tube, average channel noise hit rates below 30~Hz, and average drift tube efficiency and spatial resolution of 99\% and $82~\mu\mathrm{m}$, respectively, at an effective threshold of 15 primary electrons.

}

\keywords{Particle tracking detectors (Gaseous detectors); Muon spectrometers; Front-end electronics for detector readout; Performance of High Energy Physics Detectors}

\begin{document}
\maketitle
\flushbottom

\section{Introduction}
\label{sec:intro}
\begin{figure}[htbp]
        \centering
        \includegraphics[width=1.0\textwidth]{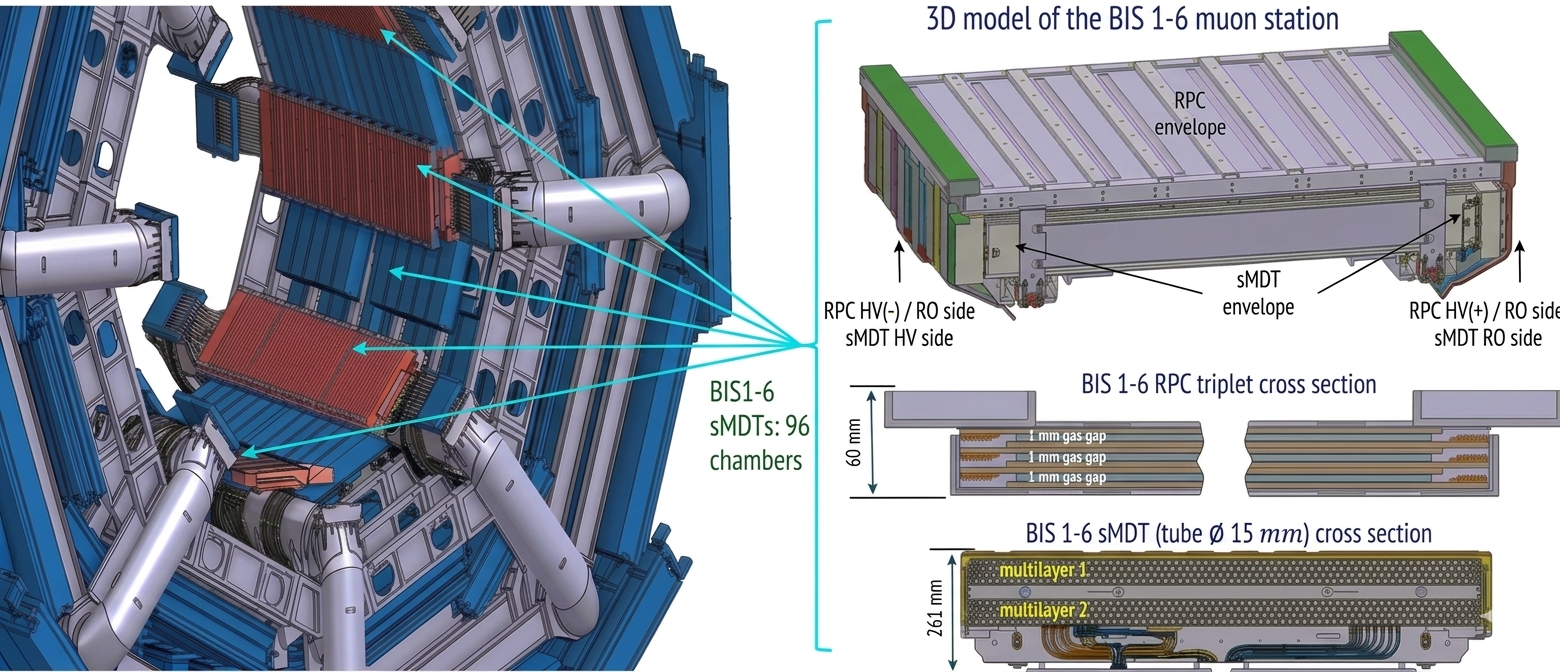}
    \caption{Schematic design of the innermost BIS1--6 barrel layer for the ATLAS Muon Spectrometer upgrade. Left: Layout of the inner barrel layer highlighting the BIS1--6 chambers. Right: 3D model and cross-section of a BIS1--6 module, combining an sMDT chamber (two multilayers of 15-mm drift tubes) with a thin-gap RPC triplet (three 1-mm gas gaps) within a 260-mm radial envelope~\cite{ATLAS-TDR-026}.}
    \label{concept_BIS16_module}
\end{figure}

The upgrade of the ATLAS Muon Spectrometer for operation at the High-Luminosity Large Hadron Collider (HL-LHC), scheduled in the Long Shutdown~3 (LS3) of the LHC from 2026 to 2030, foresees major improvements to muon triggering and tracking to ensure efficient operation at the higher luminosities~\cite{ATLAS-TDR-026}. A key ingredient is the replacement of the innermost barrel layer of Monitored Drift Tube (MDT) chambers in the small azimuthal sectors (BIS1--6 chambers) by 96 new integrated tracking and trigger modules. Each module consists of a small-diameter Muon Drift Tube (sMDT) chamber and a triplet of Resistive Plate Chambers (RPC), as shown in Figure~\ref{concept_BIS16_module}. These modules improve the first-level muon trigger acceptance and the \(p_\text{T}\) resolution, together with an order-of-magnitude increase in rate capability and detector lifetime in the high-background environment of the HL-LHC~\cite{ATLAS-TDR-026, FALSETTI2026170955, Kroha_2017}.

\if{0}
%The Phase-II upgrade of the Large Hadron Collider (LHC), planned for the 2026--2030 Long Shutdown 3 (LS3), requires substantial improvements to the ATLAS Muon Spectrometer to ensure reliable operation at the higher luminosities foreseen for the High-Luminosity LHC (HL-LHC)~\cite{ATLAS-TDR-026}. As part of this program, 96 new muon tracking and trigger modules will be installed in the innermost barrel layer (BIS1--6), replacing the existing Monitored Drift Tube (MDT) chambers, as shown in Figure~\ref{concept_BIS16_module}. Each module combines a small-diameter MDT (sMDT) chamber with a triplet of Resistive Plate Chambers (RPCs) and is designed to enhance Level-1 trigger performance, spatial resolution, and rate capability in regions subject to high background radiation ~\cite{ATLAS-TDR-026, FALSETTI2026170955}.
\fi

The sMDT technology employs thin-walled aluminum drift tubes with 15~mm diameter, reducing background-induced occupancy and space-charge effects by about a factor of eight compared with the 30~mm-diameter MDT tubes~\cite{Kroha_2017}. The smaller tube diameter also reduces the chamber thickness, enabling the integration of RPC triplets with 1~mm gas gaps within the available spatial envelope~\cite{KROHA2019445}. 
%These sMDT--RPC modules will replace the existing MDT chambers in the %barrel inner small-sector (BIS1--6) station of the Muon Spectrometer.
sMDT chamber construction has been completed: 48 sMDT chambers (plus 4 spares) were built at the Max Planck Institute for Physics (MPI) in Munich for one hemisphere of the barrel (A-type chambers) between January 2020 and December 2022, and an additional 48 chambers (plus 2 spares) were produced between 2021 and 2023 for the opposite, mirrored hemisphere (C-type chambers) by the University of Michigan (UM), in collaboration with Michigan State University~\cite{KROHA2024169818, BIS1-6_2026, Amidei_2023}. The A- and C-type chambers are identical in design, except that they are mirror images of each other with respect to the middle plane perpendicular to the beam axis through the ATLAS interaction point~\cite {sMDT_ParameterBook}.
%are the two BIS1--6 sMDT chamber designs, differing only in the tube-%numbering orientation on the readout side~\cite{sMDT_ParameterBook}. I
In total, the project comprises more than 46,000 drift tubes, corresponding to approximately 74.5~km of tube and wire length and an active detector area of about 135~m$^{2}$~\cite{sMDT_ParameterBook}. 

From 2018 to 2020, 16 BIS78 sMDT chambers were constructed as a pilot project for the Phase~2 upgrade. These chambers are similar in design to the BIS1--6 chambers but are larger and have different geometries to accommodate the ends of the BIS layer, where they overlap with the small wheels of the inner muon endcap layer~\cite{ATLAS-TDR-026, KROHA2019445}. The A-side chambers were installed in 2020 for LHC Run~3 together with thin-gap RPC triplets~\cite{BIS78_ATLAS}, while the C-side chambers will be installed during LS3 together with the BIS1--6 chambers. The experience gained from the BIS78 pilot project provided valuable input for the design and implementation of the comprehensive surface commissioning and performance evaluation program presented in this paper.

The sMDT chambers consist of two multilayers (MLs), each comprising four layers of densely packed pressurized drift tubes operated at 3 bar, which are precisely assembled~\cite{BIS1-6_2026} and glued to each other and to an aluminum space frame incorporating an in-plane optical alignment monitoring system. The only 30.5~mm high spacer structure separates the two MLs. 
%and precisely aligned to satisfy the stringent mechanical envelope %requirements of the ATLAS Muon Spectrometer. 
All assembled drift tubes are 1624~mm long, while the chamber widths and channel numbers vary: the 16 BIS1 chambers have an active area of approximately 1.64~m$^{2}$ with 560 drift tubes each, while the 80 BIS2--6 chambers have an active area of approximately 1.36~m$^{2}$ with 464 drift tubes~\cite{BIS1-6_2026, Amidei_2023,sMDT_ParameterBook}.

During chamber production at both sites, stringent quality control procedures were implemented, with particular emphasis on sense-wire positioning and chamber performance. The sense wires, made of a 97:3 tungsten--rhenium (W:Re) alloy, have a diameter of 50~$\mu$m and were tensioned to $350\pm25$~g. Wire positions were measured using precision coordinate measuring machines, yielding an average deviation of better than 10~$\mu$m over the full production campaign, well within the ATLAS specification required to achieve the target momentum resolution. Each chamber also underwent mechanical integrity, gas-tightness, and electronic tests, followed by validation of the performance with cosmic rays~\cite{KROHA2024169818, BIS1-6_2026, Amidei_2023}.

\begin{figure}[!htbp]
    \centering
    \setlength{\abovecaptionskip}{5pt}%
    \subfloat[]{%
        \includegraphics[height=4.4cm]{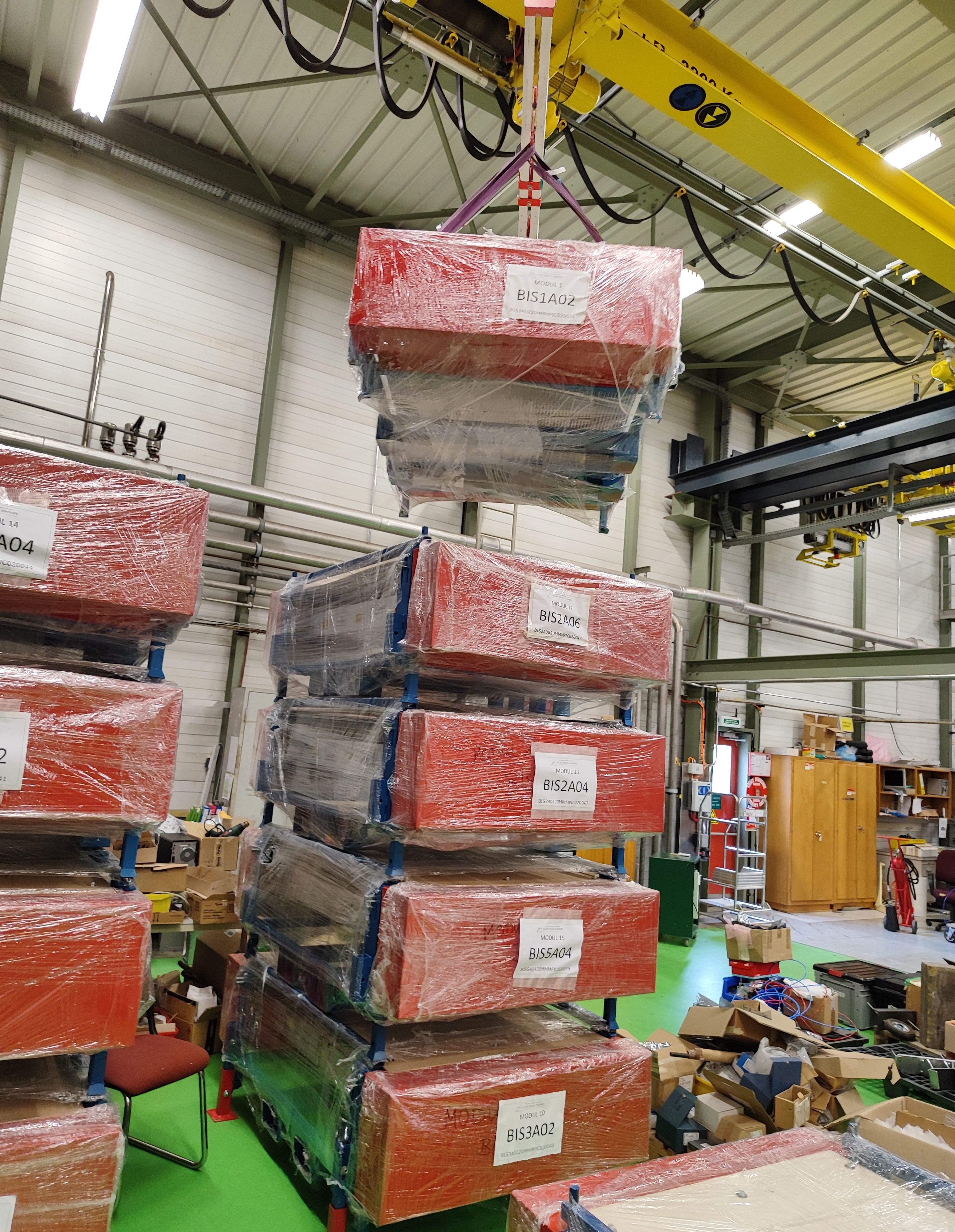}%
        \label{BB5_storage}%
    }%
    \hfill
    \subfloat[]{%
        \includegraphics[height=4.4cm]{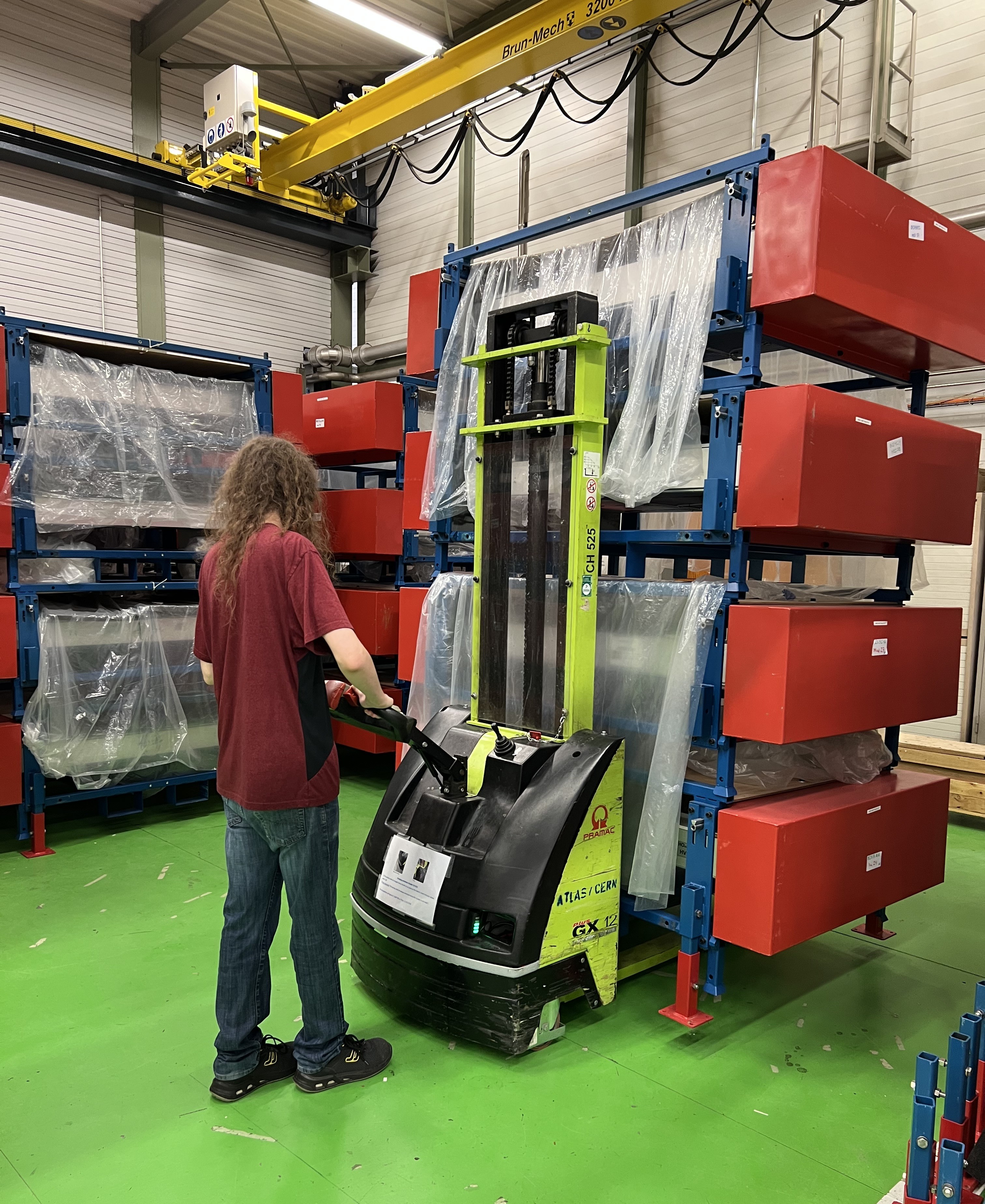}%
        \label{BB5_storage_forklift}%
    }%
    \hfill
    \subfloat[]{%
        \includegraphics[height=4.4cm]{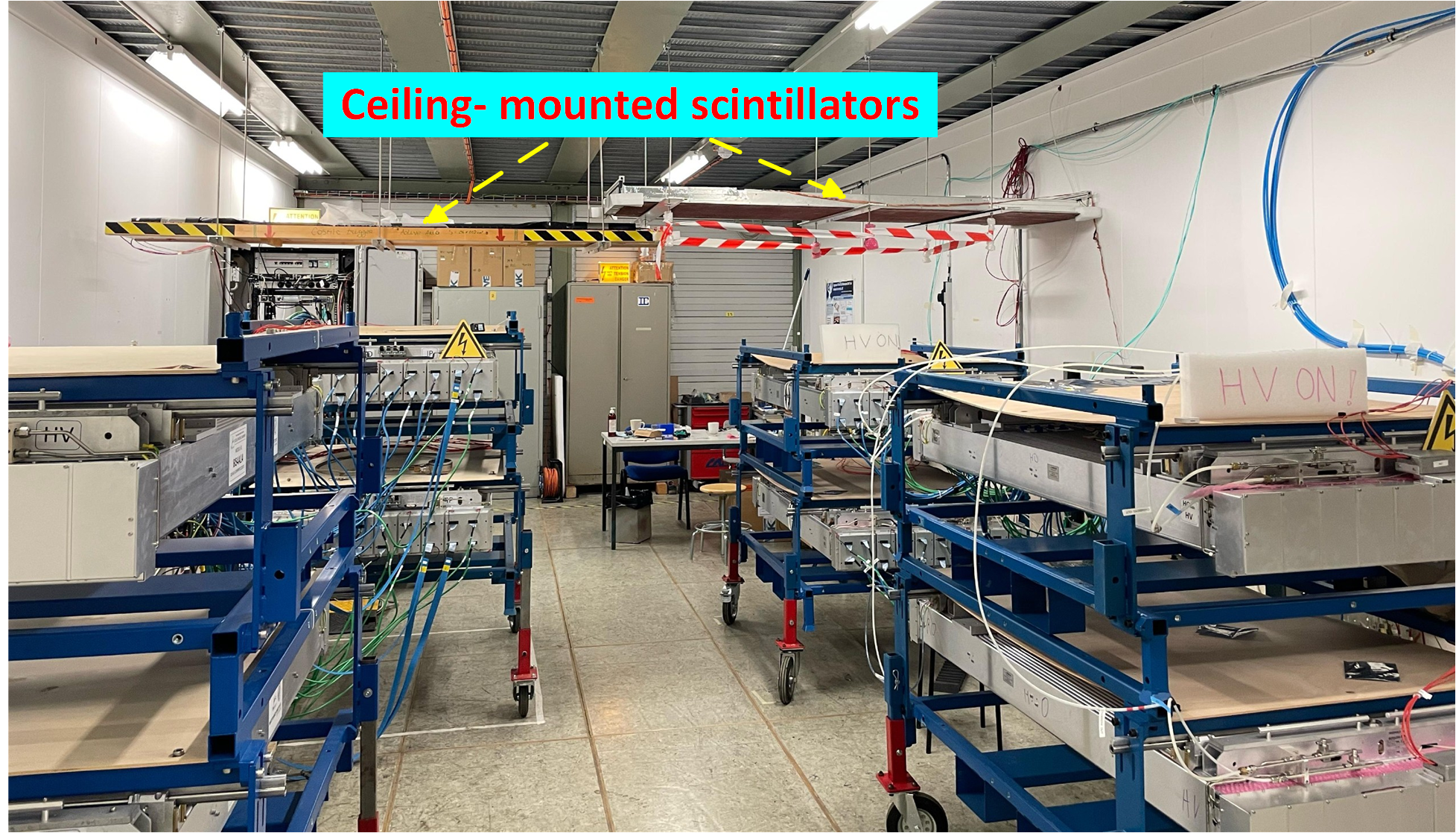}%
        \label{BB5_teststands}%
    }%
    \caption{Infrastructure for sMDT commissioning at CERN. Chambers are stored in stacks of four on transport frames and extracted using an overhead crane (a) or forklift (b). (c) Test stands for parallel commissioning of up to eight sMDT chambers, arranged in pairs at each station, with ceiling-mounted scintillators providing cosmic-ray triggers.}

    \label{BB5_storage_teststands}
\end{figure}

\begin{figure}[!htbp]
    \centering
\includegraphics[width=0.5\textwidth]{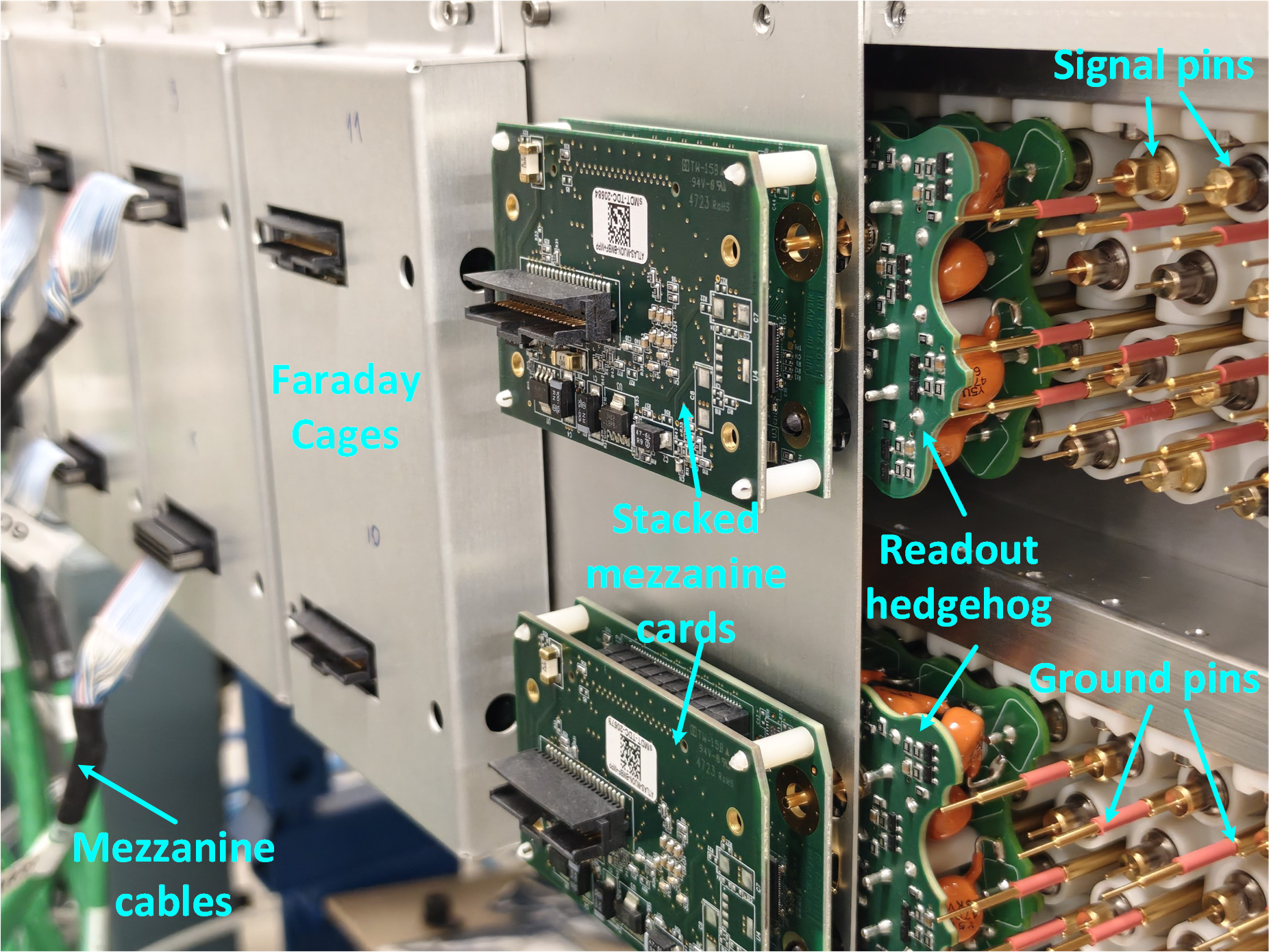} 
    \caption{Installation of the new readout hedgehog boards and mezzanine cards on an sMDT chamber. The hedgehog boards, pre-installed during chamber construction, provide signal coupling and protection while connecting each ML to \(4\times 6\) drift tubes via signal pins and to grounding screws via elongated pins between tube walls~\cite{ATLAS-TDR-026, BIS1-6_2026}. The mezzanine cards house three 8-channel ASD ASICs and a TDC ASIC, and are enclosed in an aluminum Faraday cage for shielding.}

    \label{stacked_mezzanine}
\end{figure}

By July 2024, all 102 sMDT BIS1–6 chambers from both production sites had been shipped to CERN, to a dedicated assembly and testing hall for the ATLAS muon chambers (see Figure~\ref{BB5_storage_teststands}). At this facility, the chambers undergo final validation and are equipped with the new front-end electronics required for HL-LHC operation. The new mezzanine cards contain new amplifier-shaper-discriminator (ASD) ASICs~\cite{ASD2_IEEESensor, ASD2_manual} with the same functionality as the legacy electronics~\cite{Arai_2008} but improved signal-to-noise performance and new Time-to-Digital Converter (TDC) chips~\cite{GUO2021164896} supporting continuous trigger-less readout for the MDT-based first-level muon trigger system of ATLAS at HL-LHC, both in modern 130 nm CMOS technology, mounted on the sMDT chambers on separate stacked boards for space reasons, as shown in Figure~\ref{stacked_mezzanine}. The chamber service module (CSM) concentrates the data from the on-chamber electronics for optical transmission~\cite{teng2026csm}. The detailed commissioning setup and readout scheme are described in Section~\ref{sub_sec:Readout_elex}. This paper presents the surface commissioning strategy and performance results of the BIS1--6 sMDT chambers at CERN, demonstrating their readiness for installation in ATLAS alongside the RPC triplet detectors, which are currently under construction at several sites~\cite{ATLAS-TDR-026}.

\if{0}
\fi

\section{Commissioning Procedures}
\label{sec:sMDT_commissioning}
The surface commissioning program is a key element of the BIS1--6 upgrade project to certify the readiness of the sMDT chambers for integration with the RPCs and installation in the ATLAS cavern. It consists of a standardized sequence of inspections and performance 
tests~\cite{ATLAS-TDR-026} (Table~\ref{QAQC_requirements}).
%to validate detector functionality, robustness, and long-term 
%operability under HL-LHC conditions. 
Identical procedures had already been performed at the production sites before shipment to CERN using preliminary mezzanine cards, which, however, were already equipped with the final ASD chips. The A-type chambers had also been fully commissioned again with the preliminary mezzanine cards at CERN in 2023--2024 to verify that no deterioration due to transport and long-term storage had occurred and to detect any defects as early as possible. Both noise hit rate and cosmic-ray measurements were performed at the nominal effective threshold of 20~p.e. and at a reduced threshold of 15~p.e., following the same criteria as in Table~\ref{QAQC_requirements}. 
No defects were found at this stage. 

The C-type chambers, which had been shipped in wooden boxes in sea containers across the Atlantic Ocean, had to be installed in the storage frames after they arrived at CERN.
Following retrieval from the storage area and transport to the test room (see Figure~\ref{BB5_storage_teststands}), the new mezzanine cards were installed, and the reinforcement bars for the RPC supports were pre-mounted on the sMDT chambers.
The sMDT surface commissioning campaign at CERN with final electronics started in November 2024 and was completed at the end of 2025. 
The commissioning workflow comprises six major steps, which are summarized in Table~\ref{tab:QA_QC_workflow} together with the corresponding acceptance criteria. Chambers that satisfy all acceptance criteria are returned to storage. They will subsequently be integrated with the RPC detectors and then installed in the ATLAS experiment. 

\begin{table}[!htbp]
\centering
\renewcommand{\arraystretch}{0.7}
\caption{Summary of the sMDT surface commissioning procedures and acceptance criteria}
\label{QAQC_requirements}
\label{tab:QA_QC_workflow}
\small
\begin{tabular}{p{2.6cm} p{6.5cm} p{5.0cm}}
\toprule
\textbf{Tasks} & \textbf{Work Procedure} & \textbf{Acceptance Criteria} \\
\midrule
%Visual inspection and mechanical preparation 
Visual inspection,\newline hardware installation 
& \textbullet~Verify mechanical integrity and envelope
\newline  \textbullet~Install RPC support reinforcement bars 
\newline \textbullet~Install Phase~2 mezzanine cards & 
\textbullet~No visible damage 
\newline \textbullet~All components installed\\
\midrule
%Gas preparation and tightness verification 
\hbox{Gas leak rate} measurement 
& \textbullet~Evacuate gas in chambers to below 5~mbar 
\newline \textbullet~Fill with nominal gas mixture at 3 bar 
\newline \textbullet~Measure pressure drop over at least 48 hours  & \textbullet~Leak rate 
$< 9.3 \times 10^{-3} \, \text{mbar·liter/s}$ \newline
\hspace*{1mm} per chamber \\
\midrule
In-plane optical\newline alignment system:& 
\textbullet~Verify functionality by comparing with \newline 
\hspace*{1mm} reference data from chamber construction & 
\textbullet~All four optical paths functional  
\newline \textbullet~Measured chamber deformations \newline 
\hspace*{1mm} compatible with reference data \\
\midrule
\hbox{Dark current} measurement  & 
\textbullet~Apply nominal HV of 2730 V
\newline \textbullet~Monitor dark current stability per tube & \textbullet~Dark current per tube $\le 2$~nA \\
\midrule
Noise hit rate test & 
\textbullet~Certify noise hit rates per tube at nominal \newline 
\hspace*{1mm} effective threshold of 20 p.e.
\newline \textbullet~Identify and repair noisy tubes&
\textbullet~Noise hit rate $< 1$~kHz per channel
\newline \textbullet~Average tube noise hit rate $< 100$~Hz\\
\midrule
\hbox{Cosmic ray test} & 

%\newline \textbullet~Threshold offset measurement and corrections
\textbullet~Collect at least 2 million events per chamber 
\newline \textbullet~Identify and repair hardware problems
\newline\textbullet~Replace faulty mezzanine cards 
\newline \textbullet~Identify and disconnect faulty tubes
\newline\textbullet~Evaluate hit efficiency and spatial resolution \newline 
\hspace*{1mm} to ensure stable operation in ATLAS & 
For all drift tubes:
\newline\textbullet~ADC spectra peak in 115--235~ns
\newline \textbullet~Uniform cosmic hit rate 
\newline\textbullet~Hit efficiency $\ge 99\%$ 
%\newline \textbullet~Spatial resolution $\le 100$ $\mu$m 
\newline Uniform average tube efficiency and spatial resolution 
across chambers\\
\bottomrule
\end{tabular}
\end{table}

All data and analysis results have been stored in the central chamber production database already used during all previous steps of the chamber construction and test~\cite{BIS1-6_2026}. The database is maintained at CERN and is connected to an online Web interface which provides the performance plots and the status of each chamber with traceability of the individual drift tubes, electronics boards and channels. The online monitoring and performance plots are automatically generated by the interface program. The data will be used for the muon reconstruction during ATLAS operation.

\if{0}
%Throughout the campaign, close coordination between the two teams was %essential to maintain consistent procedures, harmonized acceptance %thresholds, and uniform documentation of the results. This %comprehensive approach provides the QA needed for reliable long-term %operation and high-precision muon measurements required by the ATLAS %physics program at the HL-LHC.

%Once an sMDT chamber is certified, it will be mechanically integrated %with an RPC triplet to form an sMDT--RPC detector module, followed by %module-level QA tests before installation and commissioning in the %ATLAS cavern.
\fi

\subsection{Hardware preparations}
\label{sub_sec:preparation}
%

%The sMDT chambers were shipped to CERN in dedicated protective crates. Upon arrival, each chamber was mounted onto a blue transport frame for safe handling during commissioning and final installation. Once in the frames, chambers could be moved on integrated wheels, typically in pairs to the test room or in groups of up to four for storage.
\if{0}
The sMDT chambers arrived at CERN in protective crates. Upon arrival, they were mounted onto transport frames, allowing movement in pairs to test stands or in groups of four for storage.
%Note: The logistical details have been added by Hubert in lines 117-119: The C-type chambers, which had been shipped in wooden boxes in sea containers across the Atlantic Ocean, had to be installed in the storage frames after they arrived at CERN. Following retrieval from the storage area and transport to the test room 
\fi

\begin{figure}[htbp]
    \centering
\includegraphics[width=0.7\textwidth]{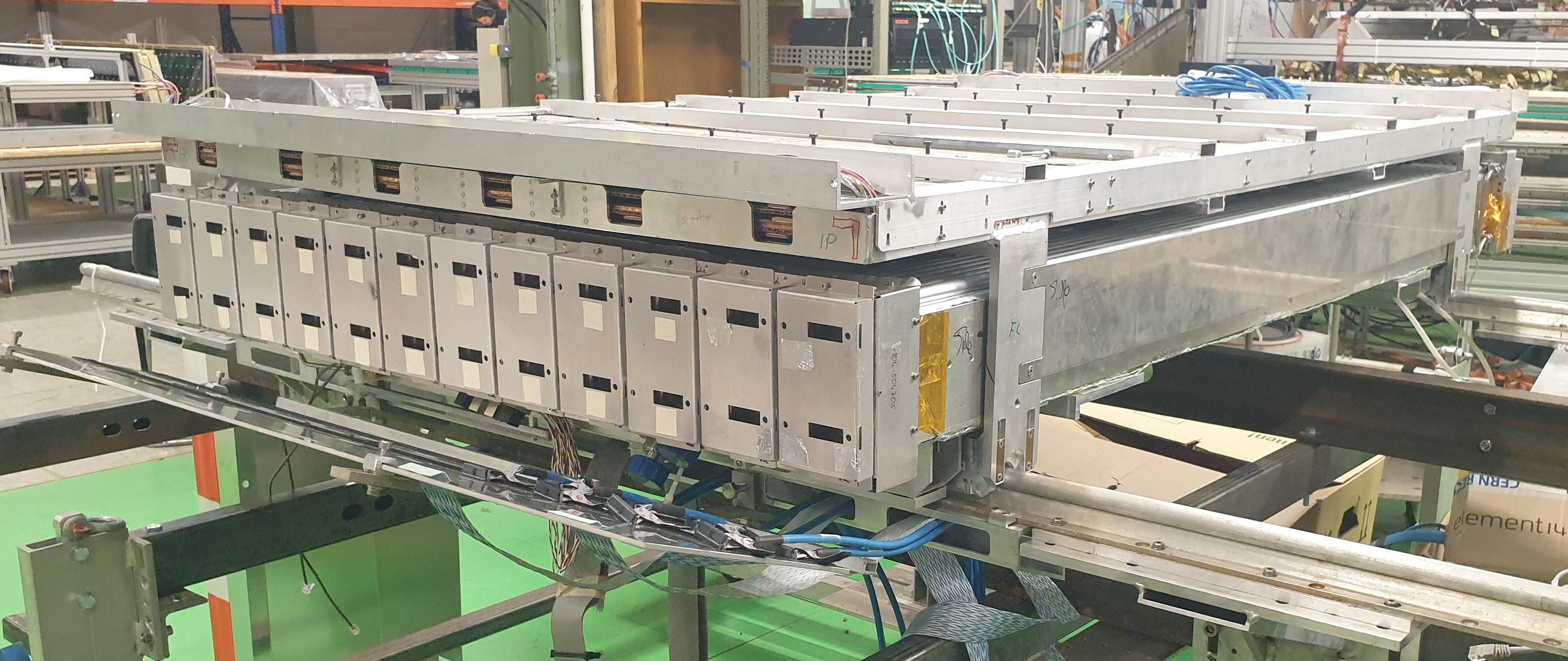}    
\caption{Prototype of an integrated BIS1 sMDT chamber with its associated RPC triplet mounted in the support frame on top, viewed from the sMDT readout side. The two detector types are mounted together on rails via interleaved bearings, as in the final ATLAS configuration. The RPC frame is supported by brackets at the four corners of the sMDT chamber, with opposite bearings along the rails connected by reinforcement bars integrated into the sMDT supports.}
    \label{fig:sMDT_RPC_integration}
\end{figure}

Each BIS sMDT chamber has to be integrated with an RPC chamber before installation in ATLAS (see Figure~\ref{fig:sMDT_RPC_integration}). The assembly and commissioning of the integrated modules at CERN is expected to start in the fall of 2026. In preparation, the reinforcement bars, which connect the pairs of rail bearings on the four RPC support brackets belonging to the same rail in order to stabilize the brackets during sliding on the rails in ATLAS, were pre-installed inside the sMDT support profiles glued to either end of the chambers. They will later be screwed to the RPC bearings. For the installation, the sMDT chambers in their transport frames were turned upside down in a special rotation stand also used later for the mechanical integration of the RPC.  

Due to envelope constraints in the ATLAS cavern, on-chamber high-voltage (HV) distribution boxes on certain chambers had to be repositioned along the HV-side Faraday cage (FC), which also required rerouting of the HV cables inside the FC. 
All on-chamber HV distribution boxes were inspected. Exposed HV connections inside the boxes may be susceptible to air discharges that could cause permanent damage. To mitigate this risk, small amounts of HV-insulating glue were applied for reinforcement and to improve long-term reliability. Additional grounding wires were installed between the circuit board inside the HV boxes and the enclosure.
Finally, before transfer to the teststands, the final Phase-2 mezzanine cards were installed on the chambers together with their individual FC boxes. The serial numbers of the installed mezzanine cards on each chamber are recorded in the database, linking them to the database of the preceding mezzanine card tests and enabling precise configuration tracking.

\subsection{Commissioning setup and readout electronics}
\label{sub_sec:Readout_elex}
The test setup for the sMDT surface commissioning at CERN is shown in
Figure~\ref{BB5_teststands}. 
The chambers are installed there in two teststands (for A- and for C-type chambers) consisting of two stacks of two chambers, allowing two teams to test up to four chambers each simultaneously. For cosmic-ray measurements, each teststand uses a large scintillation counter mounted on the room ceiling transverse to the tubes to provide the trigger for the two chambers below with coincident signals from photomultipliers at both ends.
\if[0]
\begin{figure}[!htbp]
    \centering
    \subfloat[]{
        \includegraphics[width=0.32\textwidth,height=3.5cm]{Figures/sMDT_Commissioning_TestRoom.jpg}
        \label{fig:sMDT_TestRoom}
    }
    \hspace{-0.3cm} 
    \subfloat[]{
        \includegraphics[width=0.32\textwidth,height=3.5cm]{Figures/Teststand_UM.jpg}
        \label{fig:Teststand_UM}
    }
    \hspace{-0.3cm} 
    \subfloat[]{
        \includegraphics[width=0.32\textwidth,height=3.5cm]{Figures/Teststand_MPI.jpg}
        \label{fig:Teststand_MPI}
    }    
\caption{(a) The commissioning setup in a shared test room at CERN equipped with power and trigger systems and capable of testing up to eight chambers in parallel by two teams; (b) UM test stand; and (c) MPI test stand.}
    \label{fig:workspace}
\end{figure}
\fi
One end of each sMDT chamber is connected to the HV power supply through 24-channel HV hedgehog boards which distribute the HV to all drift tubes via daisy-chained connections within each tube layer. On the opposite end, readout hedgehog boards connect the tubes to the front-end electronics on stacked mezzanine cards, providing AC coupling and input protection for the readout ASICs~\cite{BIS1-6_2026, Amidei_2023} as well as ground connections to the drift tube walls (see Figure~\ref{stacked_mezzanine}). The hedgehog boards have been pre-installed and enclosed in the FCs at the chamber construction sites. The mezzanine cards are connected to the readout hedgehog boards through slits in the FC cover plates and are themselves enclosed in individual FC boxes (see Figure~\ref{stacked_mezzanine}).

\begin{figure}[htbp]
    \centering
    \includegraphics[width=1.0\textwidth,height=4.8cm]{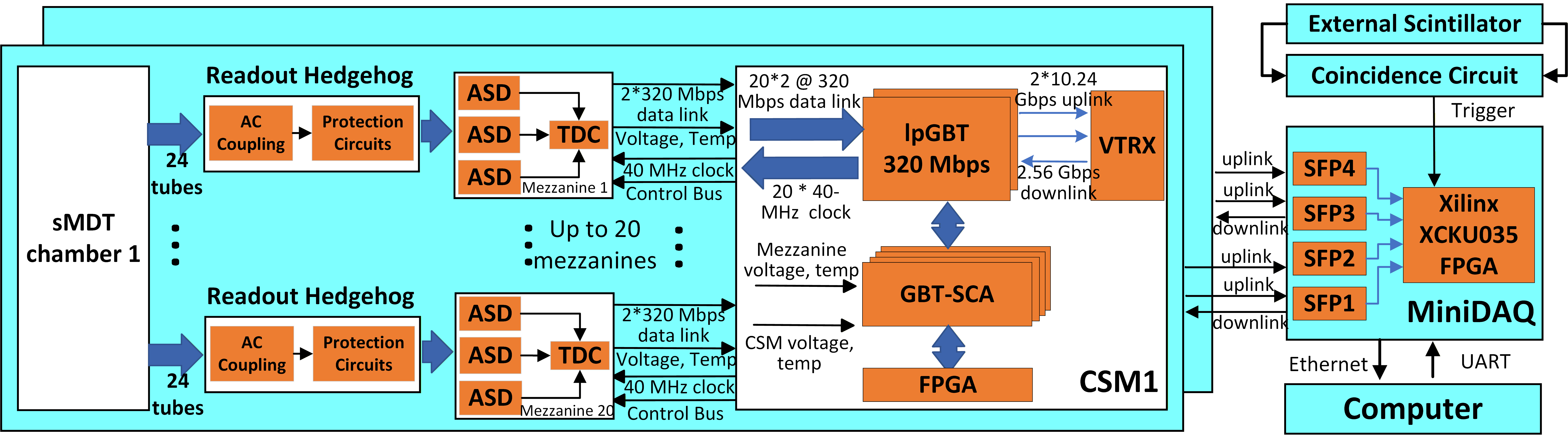}

    \caption{Schematic of the Phase~2 sMDT chamber readout electronics chain with the MiniDAQ system developed for detector and electronics commissioning.}
    \label{fig:readout_scheme}
\end{figure}

\begin{figure}[htbp]
    \centering
    \subfloat[]{
        \includegraphics[width=0.45\textwidth]{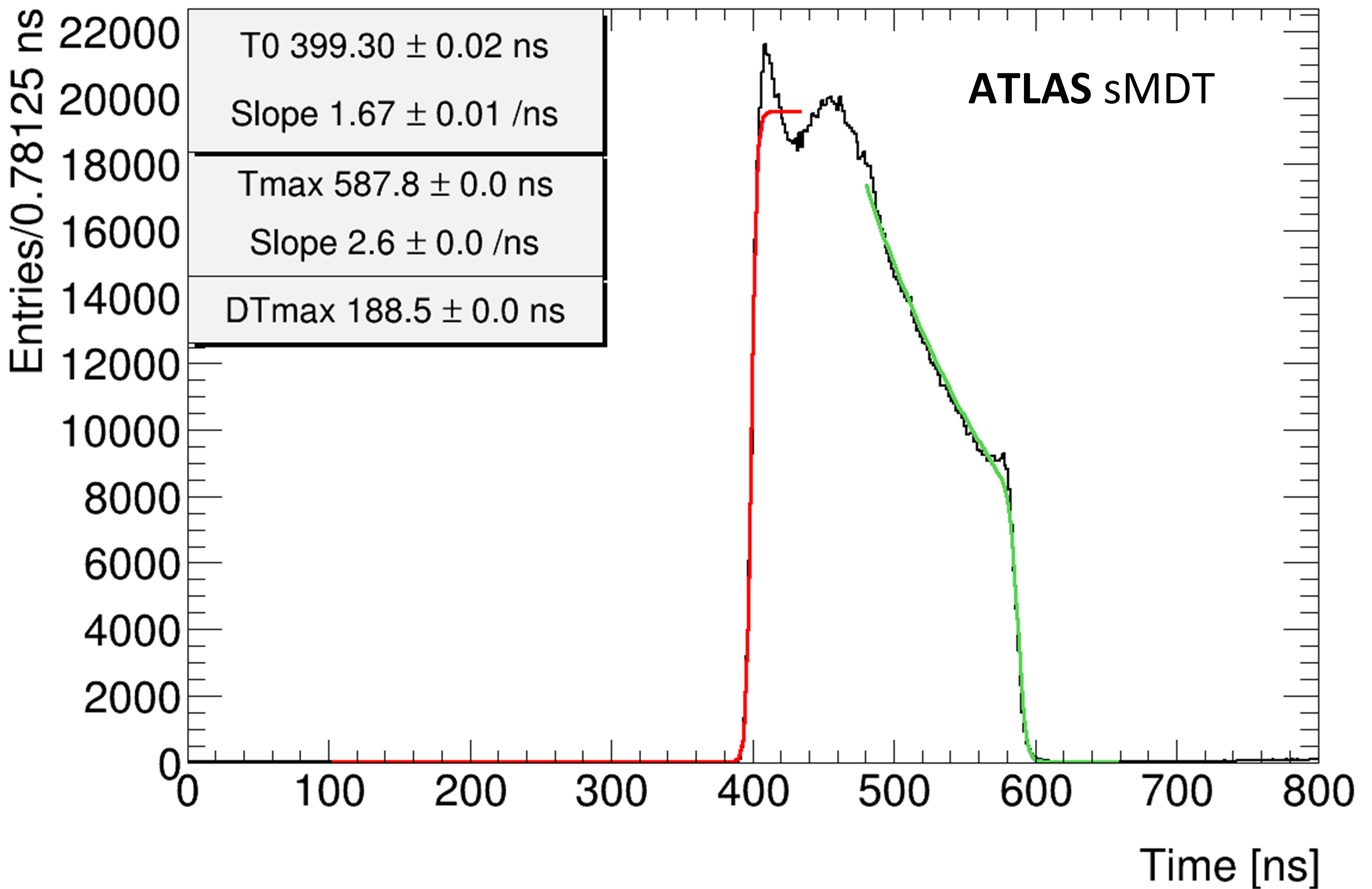}
        \label{fig:spectra_tdc}
    }
    \subfloat[]{
        \includegraphics[width=0.45\textwidth]{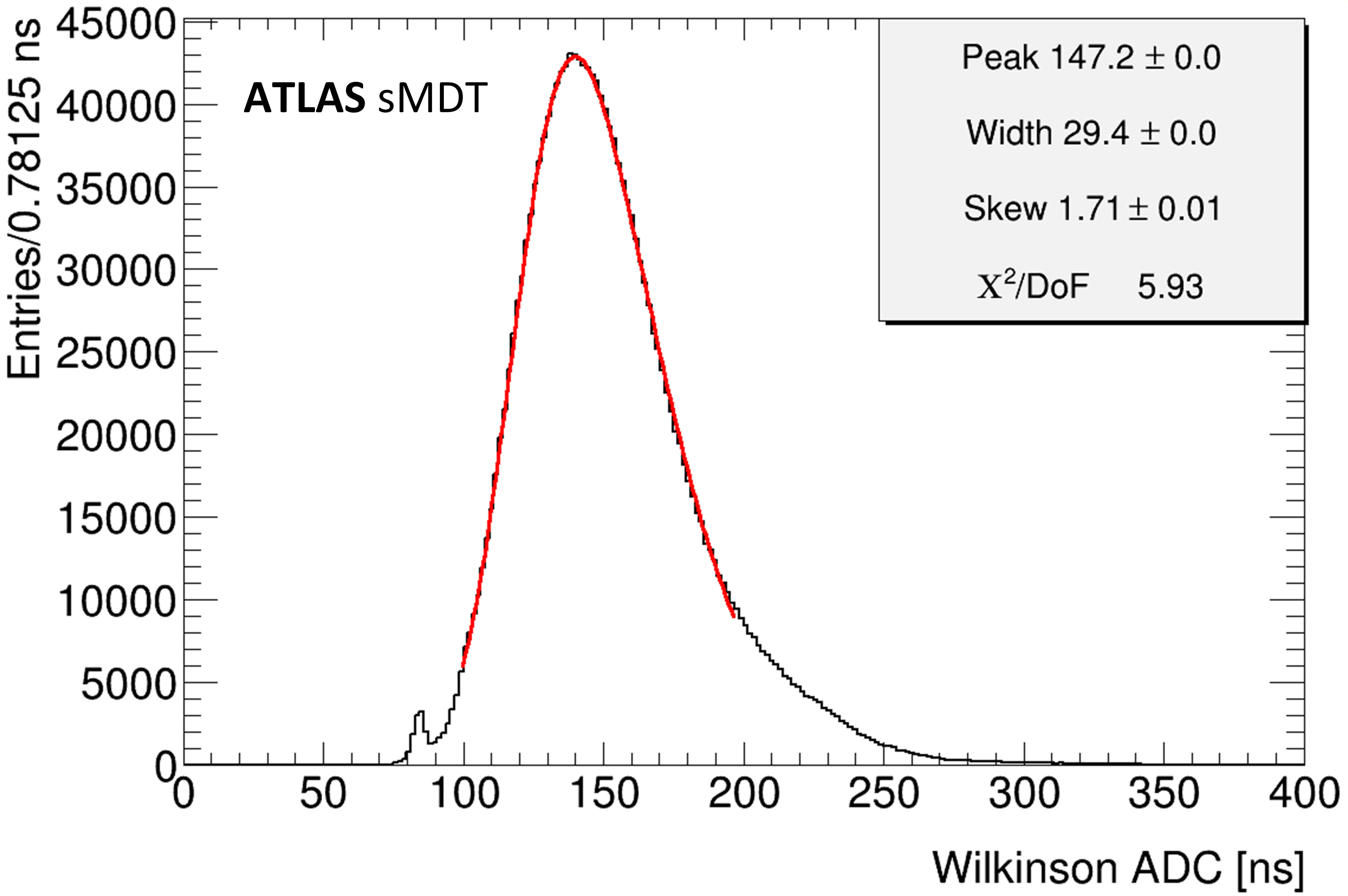}
        \label{fig:spectra_adc}
    }

    \caption{Representative drift time (a) and ADC (b) spectra of an sMDT drift tube from a cosmic ray test, under nominal operating conditions with Ar:CO\(_2\) (93:7) at 3~bar and 2730~V HV.}
    \label{fig:spectra_adc_tdc}
\end{figure}

The sMDT readout scheme used for the commissioning is shown in Figure~\ref{fig:readout_scheme}. The on-chamber readout electronics on the mezzanine cards~\cite{ASD2_manual, GUO2021164896} have been described in Section~\ref{sec:intro}.
The ASD chips amplify and shape the drift tube signals, perform signal charge-to-time conversion via a Wilkinson-type Analog-to-Digital Converter (WADC) that generates an output pulse with a width proportional to the signal amplitude, and provide signal discrimination above programmable threshold settings~\cite{ASD2_IEEESensor, ASD2_manual}. 
%The ADC circuit on the ASD chips digitizes the charge in the leading edge of the shaped signal as measure of the signal amplitude. 
The TDC chip then digitizes the leading edge of the arrival time (ToA) and the trailing edge of the discriminated signal, with a bin width of 0.78~ns~\cite{GUO2021164896}. 
The ToA distribution relative to the trigger time corresponds to the drift-time spectrum. From the drift time (t), the drift radius (r) is derived based on the space-to-drift-time (r-t) relation determined for the test of each chamber using an auto-calibration algorithm. A representative drift-time spectrum from a cosmic-ray test is shown in Figure~\ref{fig:spectra_tdc}, exhibiting a rise time and a drift-time range of approximately 190~ns, consistent with the nominal characteristics of the 15-mm diameter sMDT tubes under nominal operating conditions with Ar:CO\(_2\) (93:7) at 3~bar and 2730~V HV~\cite{Kroha_2017}. The drift gas temperature and pressure are continuously monitored. Their variations are taken into account in the auto-calibration procedure of the r-t relation.

The corresponding pulse length (the “ADC reading”) is monotonically related to the charge, contained in the rising edge of the analog input signal, which is a measure of its slope and the basis for a later correction of the slewing effect (Section~\ref{sec:results} and Eq.~\ref{eq:TS_corr}). The ADC spectrum from a cosmic-ray test is shown in Figure~\ref{fig:spectra_adc}, exhibiting a signal peak around 150~ns, well separated from the noise below 100~ns, with the WADC run-down current set to 3.8~$\mu$A (digital code 2), which was used consistently throughout the sMDT commissioning tests~\cite{ASD2_manual}.

%Representative ADC and drift-time ("TDC") spectra from a cosmic-ray test are shown in Figures~\ref{fig:spectra_adc} and~\ref{fig:spectra_tdc}, respectively. The ADC spectrum shows a peak around 150~ns, well separated from the noise charge level below channel 100~ns, for the WADC run-down current of 3.8~$\mu$A (digital code is 2) used throughout the sMDT commissioning tests~\cite{ASD2_manual}. The TDC spectrum exhibits a rise time and a drift-time range of approximately 190~ns, consistent with the nominal characteristics of the 15-mm diameter sMDT tubes under the operating conditions during the commissioning~\cite{Kroha_2017}. The drift gas temperature and pressure are continuously monitored. Their variations are taken into account in the auto-calibration procedure of the r-t relation.%

The noise hit rate and spatial resolution of the drift tubes depends on the discriminator threshold of the ASD chip. The threshold is set by an 8-bit code ranging between 0 and 255, where the discriminator thresholds for negative signal charges correspond to the lower half of the codes, with zero at 127.5 and increasing with decreasing code values. The conversion of the code numbers $N_{\text{code}}$ into signal amplitude 
$A_{\text{thr}}$ is given in 3.3~mV steps by the relation 
$A_{\text{thr}}=3.3~mV\cdot (N_{\text{code}}-127.5)$~\cite{ASD2_manual}. 
In addition, hysteresis in the threshold crossing of signals can be set by a 4-bit code ranging from 0 to 15 with the aim of suppressing multiple threshold crossings due to noise hits. 
It has been demonstrated in this work (see Section~\ref{sec:results}), and confirmed in \cite{Kortner_threshold_calibration}, that increasing hysteresis effectively increases the discriminator threshold and thus the drift tube spatial resolution. 

The discriminator thresholds had been determined in muon test-beam measurements in terms of numbers of primary ionization electrons (p.e.) for the nominal threshold code setting of 114 and hysteresis settings of 7 and 14~\cite{Kortner_threshold_calibration}. These settings were adopted as the starting point for the surface commissioning at CERN. Threshold/hysteresis settings 114/14 and 108/7 correspond to the effective threshold of 20 p.e., as well as settings 120/14 and 114/7, corresponding to 15 p.e. The first code combinations for each effective threshold have been used for A-type and the second for C-type chambers in the commissioning campaign for practical reasons.

\if{0}
%%%%%%%%%%%%%
%The sMDT readout scheme employed during testing is shown in Figure~\ref{fig:readout_scheme}. During commissioning, chambers are equipped with final-production sMDT mezzanine cards. The new cards feature upgraded Amplifier-Shaper-Discriminator (ASD) and Time-to-Digital Converter (TDC) chips on separate stacked boards to fit the tighter BIS chamber envelope~\cite{Arai_2008, ATLAS-TDR-026, ABOVYAN2019374, GUO2021164896}. The ASD chip performs signal amplification, shaping, Wilkinson-type conversion, and discrimination~\cite{ASD2_IEEESensor}. Following discrimination of the Wilkinson-converted signal, the TDC measures both the leading-edge arrival time (TOA) and the time-over-threshold (ToT)~\cite{GUO2021164896}. The ADC (Analog-to-Digital Converter) spectrum thus represents the ToT distribution, which is monotonically related to the deposited charge and reflects the energy-loss spectrum. The TDC spectrum (i.e., the TOA distribution) provides the drift-time information, from which the drift radius and spatial resolution are derived. Typical ADC and TDC spectra of a cosmic run are shown in Figure~\ref{fig:spectra_adc} and Figure~\ref{fig:spectra_tdc}. In the ADC spectrum, the signal peak is observed around channel 160, while the noise peak appears below channel 100. The TDC spectrum exhibits a drift time interval of approximately 190~ns for the sMDT chambers, consistent with the expected drift-time characteristic of the 15-mm-diameter tubes under the chamber operation environment at BB5.
\fi

The data from the front-end readout cards are transmitted to the chamber service module (CSM) that connects up to 20 mezzanine cards via shielded signal cables (see Figure~\ref{stacked_mezzanine})~\cite{ATLAS-TDR-026, teng2026csm}. To accommodate the higher hit rates expected at the HL-LHC, the new CSM processes 40 data channels at 320~Mbps per channel and multiplexes the data to two optical uplinks operating at 10.24~Gbps, each using two low-power Gigabit Transceiver (lpGBT) chips~\cite{lpGBT, GUO2023167671}. A separate 2.56~Gbps optical downlink distributes the clock, control, and configuration signals to the mezzanine cards. The CSM also features four GBT-SCA (Gigabit Transceiver—Slow Control Adapter) chips for monitoring the voltages and temperatures of both the mezzanines and the CSM itself~\cite{gbt_sca}. Monitoring of the mezzanine cards inside the FC showed typical operating temperatures of about 40\(^\circ\)C, with an supply voltage of 3.6~V and current of about 570~mA.

The Mini Data Acquisition (MiniDAQ) system can read out two CSMs simultaneously, perform trigger matching, and transfer the events to the data storage and analysis computer~\cite{GUO2021164896}. BIS1 chambers carry 24 mezzanine cards and thus two CSMs, while BIS2--6 chambers contain 20 mezzanine cards and one CSM. Each mezzanine card reads out 24 tubes, except for the cards at the chamber edges, which read out only 20 tubes~\cite{sMDT_ParameterBook}. Thus, one MiniDAQ system can read out either two BIS2--6 chambers or one BIS1 chamber. Identical MiniDAQ setups were deployed for the two cosmic ray teststands.
An online graphical user interface (GUI) runs on the DAQ computer to record data from the miniDAQ system. It provides real-time visualization of ADC and TDC spectra, channel hit rates, and error messages, enabling immediate identification of anomalous detector behavior or hardware issues during data taking.
%These online monitoring capabilities proved essential for rapid debugging and efficient quality assurance %throughout the sMDT surface commissioning campaign.

\subsection{Gas tightness certification}
\label{sub_sec:gas_leak}
The first surface commissioning test was to re-certify the gas tightness of the chambers after the transport to CERN.
A maximum allowed gas leak rate of a (s)MDT chamber is specified~\cite{ATLAS-TDR-026} to
$10^{-5}~\mathrm{mbar \cdot liter \cdot s^{-1}}\times 2N_{\mathrm{tube}}$
corresponding to 
$9.3 \times 10^{-3}~\mathrm{mbar \cdot liter \cdot s^{-1}}$ for BIS2--6 chambers with $N_{\mathrm{tube}}=464$ and  $11.2\times 10^{-3}~\mathrm{mbar \cdot liter \cdot s^{-1}}$ for BIS1 chambers with $N_{\mathrm{tube}}=560$, respectively.
After the installation of the hardware components, the chambers were transferred to the teststands, evacuated to below \SI{5}{mbar} absolute, and filled with the nominal Ar:CO\textsubscript{2} (93:7) gas mixture at \SI{3}{bar} absolute. It was then disconnected from the gas supply by closing the on-chamber gas valves.

After at least \SI{1}{hour}, an initial pressure reading $P_i$ was taken at a gas temperature $T_i$ using high-accuracy pressure gauges
%\footnote{OMEGA DPG210-100A} 
connected to each ML. The pressure was monitored together with timestamps and temperature readings from the 12 temperature sensors distributed across the chamber surface. A second pressure measurement $P_f$ at temperature $T_f$ was taken at least 48 hours later to determine the gas leak rate $Q$ according to the relation 

\begin{equation}
Q = \frac{\Delta P_{\text{corr}} \cdot V}{\Delta t} \quad \left[ \text{mbar} \cdot \text{liter} \cdot \text{s}^{-1} \right]
\label{eq:leak_rate}
\end{equation}
where $V$ is the chamber gas volume (284 liters for BIS1 and 235 liters for BIS2--6 chambers) and 
\begin{equation}
\Delta P_{\text{corr}} = P_f \left( \frac{T_{\text{ref}}}{T_f} \right) - P_i \left( \frac{T_{\text{ref}}}{T_i} \right)
\label{eq:leak_rate_cal}
\end{equation}
the temperature-corrected pressure drop with reference temperature $T-293.15$~K.
The extended measurement interval $\Delta t$ reduces the measurement uncertainty.

Chambers with gas leak rates above the acceptance limit underwent further investigation and repair. Leaks were localized by introducing a small amount of helium into the chamber and scanning gas connections and O-ring seals with a high-sensitivity helium leak detector.
%\footnote{VeeCo MS-40B}. 
Identified leaks were repaired by replacing the affected O-ring or gas connector. The chamber was then evacuated, refilled, and tested again. The procedure was repeated until the leak rate requirement was achieved.
In a few cases, leaks were found in individual tube walls and repaired by applying a small amount of epoxy glue certified against outgassing.
An unsuccessful repair would have required permanently disconnecting the tube from the gas system and classifying it as a "dead tube". No such failures occurred during the commissioning campaign. 

\begin{figure}[htbp]
    \centering
    \includegraphics[width=0.6\textwidth]{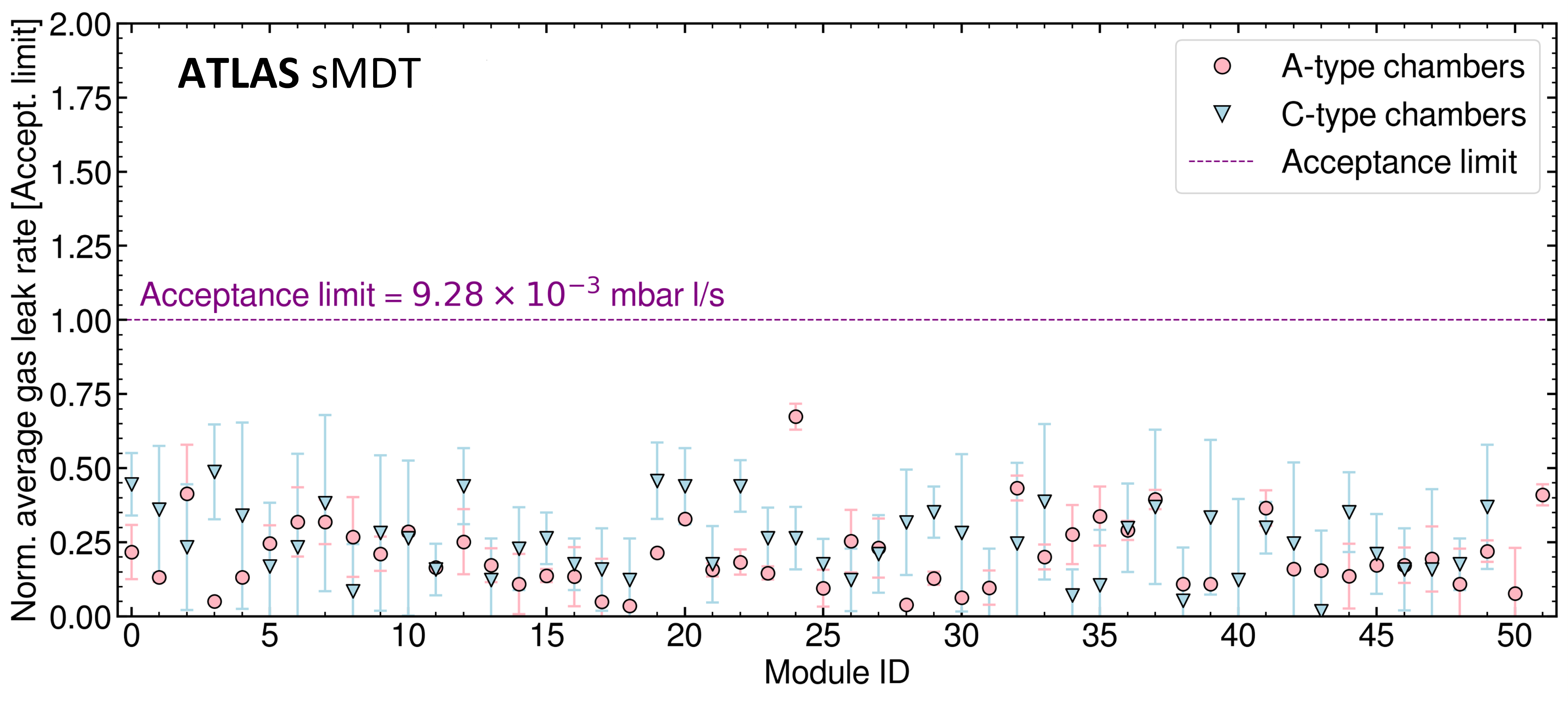}
\caption{Measured gas leak rates at the nominal operating gas mixture and pressure, normalized to the acceptance limit indicated by the dashed line, for all A- and C-type chambers at CERN. Module ID reflects the production order at each site.} 
    \label{fig:LeakRate_results}
\end{figure}

Figure~\ref{fig:LeakRate_results} shows the resulting gas leak rates for all A- and C-type BIS1--6 sMDT chambers, normalized to the acceptance limit for BIS2--6 chambers. As before at the production sites~\cite{BIS1-6_2026, Amidei_2023}, all chambers at CERN meet the strict requirement.  
On average, they exceed it by a factor of 5. BIS1 sMDT chambers contain 5600 O-ring\footnote{EPDM O-ring} seals, BIS2--6 chambers 4640, 10 per tube.

\subsection{Test of the optical planarity monitoring system}
\label{sub_sec:alignment}
The optical
%\footnote{Red Alignment System of NIKHEF} 
in-plane alignment system is integrated into the spacer frame between the two MLs~\cite{ATLAS-TDR-026, Amidei_2023} to monitor the planarity of the chambers, as in the legacy MDT chambers, in particular the torsion angle between the readout and HV ends. It consists of two longitudinal and two diagonal RASNIK straightness sensors~\cite{Beker_2019}, connected to light paths with two infrared light-emitting diodes (LEDs) and two pixel sensor cameras mounted on opposite chamber ends at the four corners of the spacer frame. For each light path, the pattern of a coded mask illuminated by an LED is imaged onto the opposite camera through a lens mounted in the middle between the LED and camera. Analysis of the projected images measures deviations from straightness of 
each path, defined by reference readings taken on the chamber precision assembly jigging with a precision of approximately 1~\textmu m.

The in-plane alignment system was validated during sMDT surface commissioning by comparing the readings of all RASNIK sensors of a chamber with the reference readings. For all chambers, the observed deviations, including torsion, were within small ranges, consistent with the mounting of the chambers on the rails of the transport frames on three-point supports. 
In seven chambers, partial failure of cameras or LEDs reduced the number of functional optical paths from four to two, which means, in particular, the loss of one diagonal path and thus the torsion measurement. The RASNIK sensor elements cannot be replaced since this would imply the loss of the reference measurement. One A-type chamber could be replaced by a suitable spare chamber.

The chamber planarity information can be recovered for isolated chambers by the global planarity monitoring systems of the chamber layers in each azimuthal sector of the muon spectrometer. For the global alignment within and between layers of a sector, the chambers carry platforms for mounting additional optical sensors, which have been glued to the ML at the bottom in Figure~\ref{concept_BIS16_module} with high precision with respect to the sense wire grid during chamber assembly~\cite{BIS1-6_2026}. The global alignment sensors will be recovered from the legacy BIS1-6 MDT chambers after their deinstallation in LS3 and subsequently installed on the alignment sensor platforms of the new sMDT chambers.

\subsection{Threshold offset measurements}
\label{sub_sec:thr-offset}

\begin{figure}[htbp]
    \centering
    \hskip -2mm
    \subfloat[]{
        \includegraphics[width=0.31\textwidth]{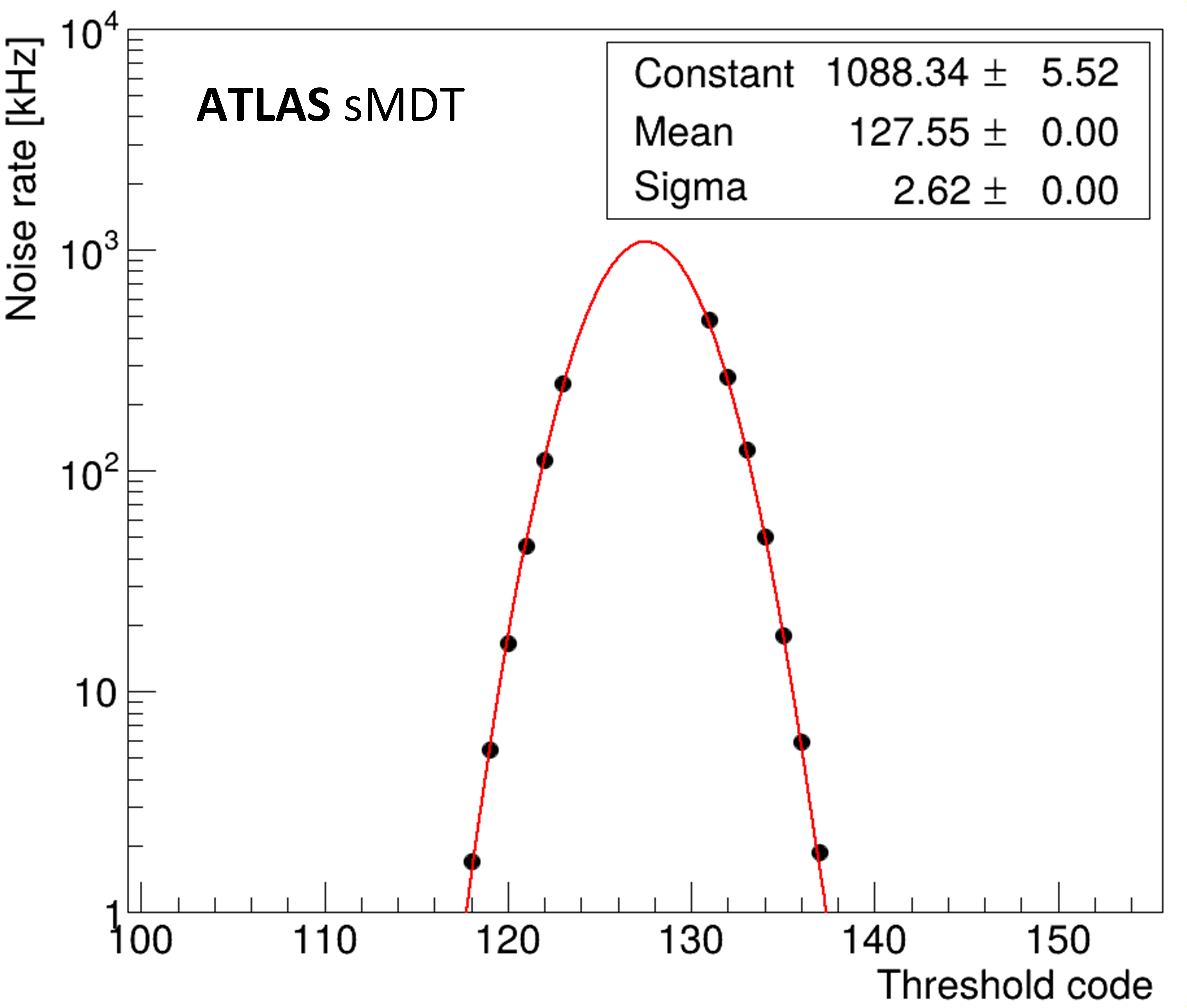}
        \label{fig:threshold_scan_gaussian}
    } 
 %   \hskip -3mm
    \subfloat[]{
        \includegraphics[width=0.32\textwidth]{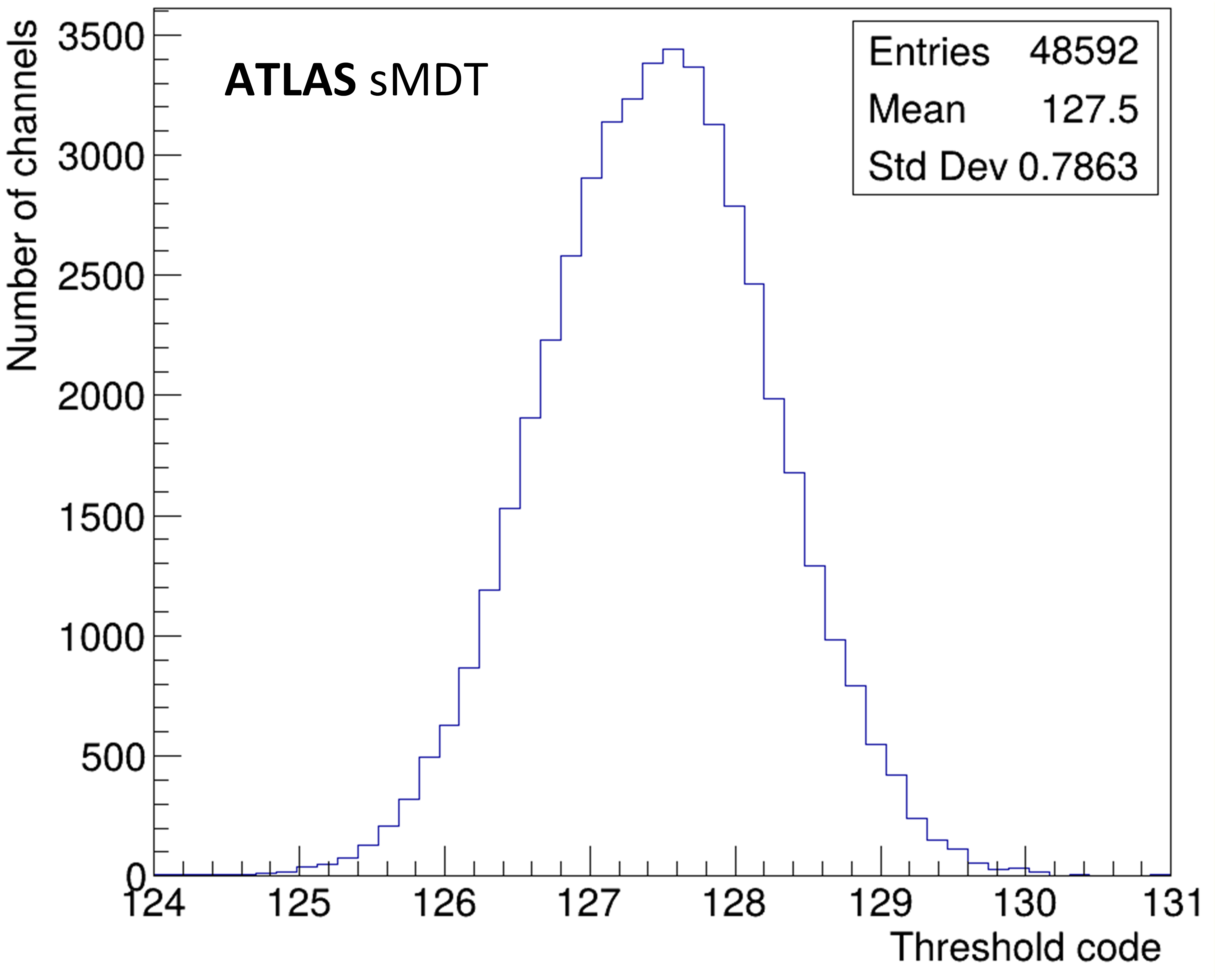}
        \label{fig:threshold_offset_distribution}
    }
%    \hskip -3mm
    \subfloat[]{
        \includegraphics[width=0.32\textwidth]{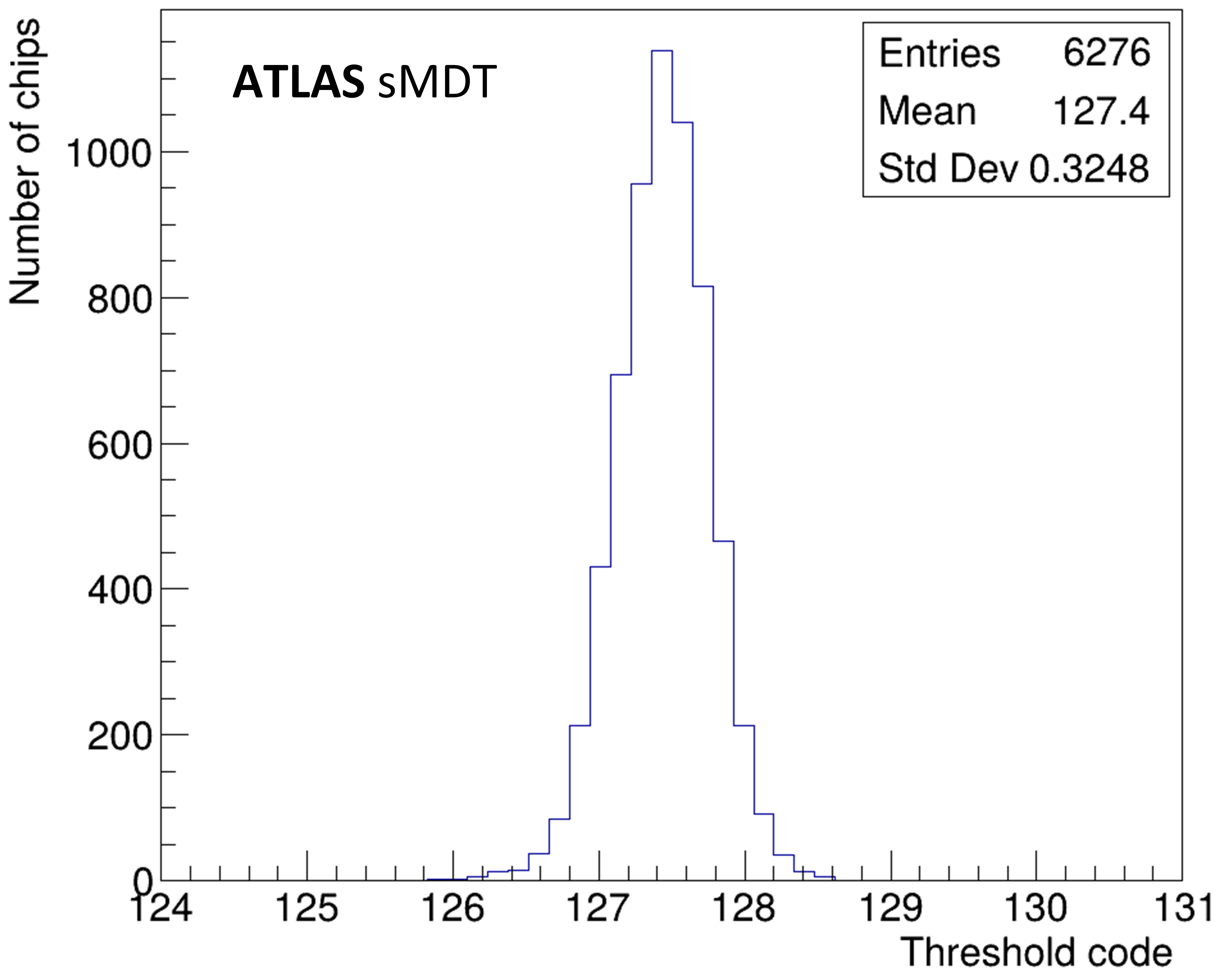}
        \label{fig:threshold_offset_chiptochip}
    }    
\caption{(a) Noise hit rate versus ASD threshold setting for an example channel with HV off. The threshold offset (peak position minus zero setting of 127.5) is only 0.05 in this case. (b) Threshold offset distribution for all channels of the BIS1-6 sMDT chambers, and (c) distribution of the average offsets of the channels of all ASD chips with even smaller variations.}
    \label{fig:threshold_scan}
\end{figure}

To determine the threshold offset corrections for each channel with respect to the nominal set discriminator threshold (see Section~\ref{sub_sec:Readout_elex}), which are caused by variations in the ASD chip production, 
threshold scans were performed measuring the noise hit rate as a function of the threshold setting for fixed hysteresis settings and with HV turned off.
%(for HV on, additional cosmic muon counts contribute, which do not %affect the result although comparable in rate at the nominal threshold). 
The noise hits fluctuate randomly around the baseline with positive and negative amplitudes with a Gaussian distribution such that the noise hit rate is expected to peak at the threshold setting 127.5, corresponding to zero discriminator threshold. 
%The noise rates were measured as a function of the main %threshold, and the resulting curves were used to determine the offset stored in the database. 
%These measurements provide the offset corrections needed for %threshold compensation and support noise-performance evaluation %and threshold optimization.
Offsets from that peak position for all individual channels of A- and C-type chambers were determined from Gaussian fits to the scan curves (see an example in Figure~\ref{fig:threshold_scan_gaussian}) and stored in the database. Points at threshold values near zero, leading to noise hit rates above the saturation of the electronics bandwidth, were excluded from the fits.

Figure~\ref{fig:threshold_offset_distribution} shows the distribution of the threshold offsets for all channels, with an RMS value of 0.79 threshold counts. These values are considerably smaller than those of the legacy ASD chip and have negligible effects on the chamber performance. Thus, they have not been generally applied for the noise hit rate evaluation and cosmic ray measurements at the nominal threshold settings.
In Figure~\ref{fig:threshold_offset_chiptochip}, the distribution of the average threshold offsets of the individual 8-channel ASD chips is shown, which is less than half as wide as the distribution of the channel-to-channel variation, and only common corrections for all the channels of a chip can be programmed. 
%For each readout channel, the scan data were used to %obtain the noise rate as a function of the main %threshold, from which a Gaussian fit determined the %threshold offset. %Figure~\ref{fig:threshold_scan_gaussian} shows the noise %rate versus main threshold for an example channel with %HV on. 
\if{0}
The fitted mean of \SI{-0.13}{mV} gives the threshold offset for this channel. Figure~\ref{fig:threshold_offset_distribution} shows the threshold-offset distribution for all channels in an example chamber, ranging from \SI{-6}{mV} to \SI{3}{mV}, with a mean of \SI{-1.09}{mV} and an RMS of \SI{1.52}{mV}. This result is representative of the threshold-offset distributions observed across all sMDT chambers.
\fi

\subsection{Noise hit rate evaluation}
\label{sub_sec:noise_hv}
\begin{figure}[!htbp]
    \centering
    \includegraphics[ height=4.2cm]{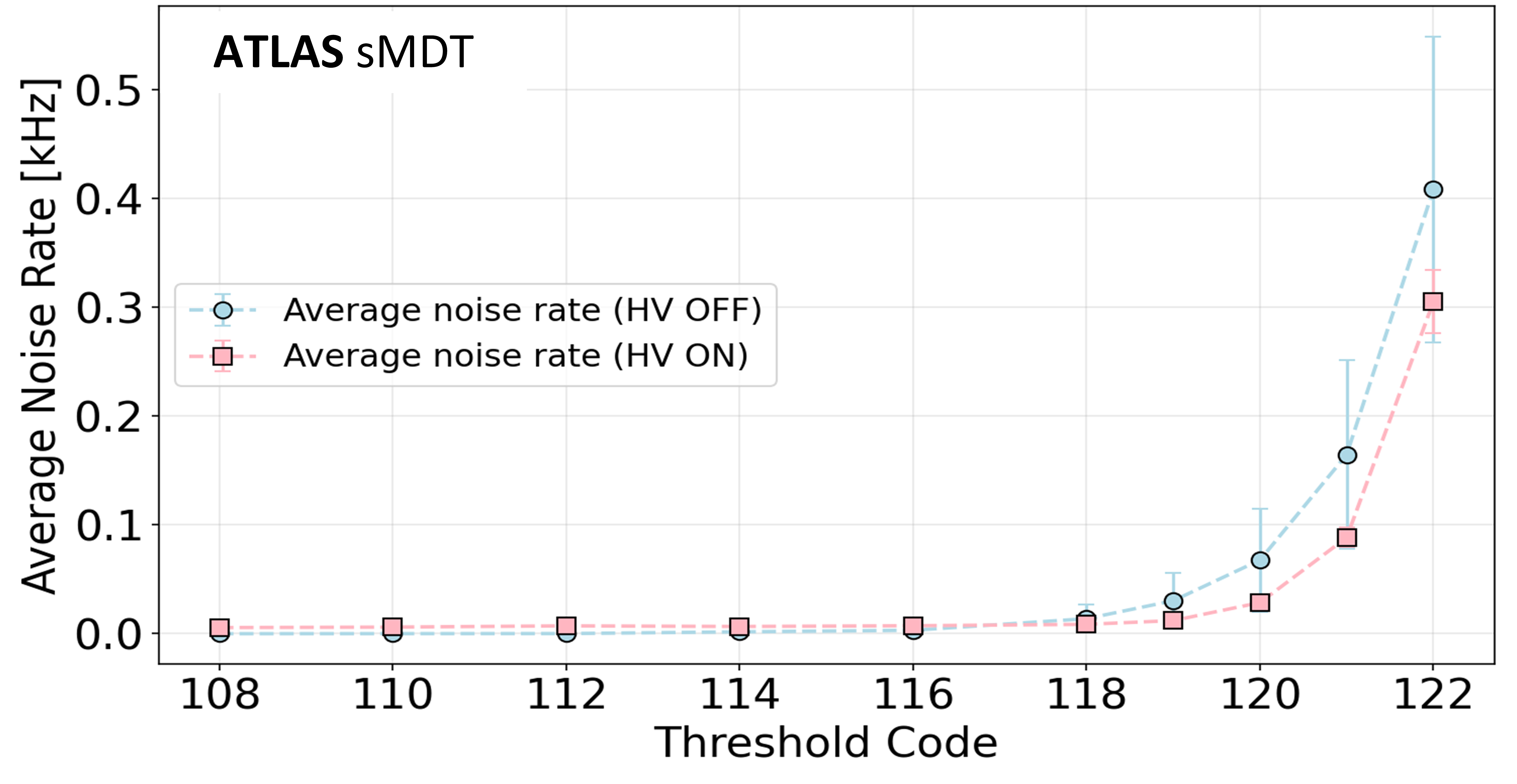}
    \caption{Average noise hit rate of the drift tubes of an A-type chamber as a function of the threshold code for hysteresis setting 14 with HV turned on and off. The exponential rise of the noise hit rate starts above code 114, where the contribution of the cosmic muon counting rate also increases.}
    \label{fig:threshold_scan_opt}
\end{figure}

\begin{figure}[!htbp]
    \centering
    \subfloat[]{   
%       \includegraphics[width=0.49\textwidth,height=4.5cm]
%       {Figures/performance_new/
%       MPI_HV_ON_vs_HV_OFF_thr_120_no_offset.pdf}
        \includegraphics[width=0.49\textwidth]{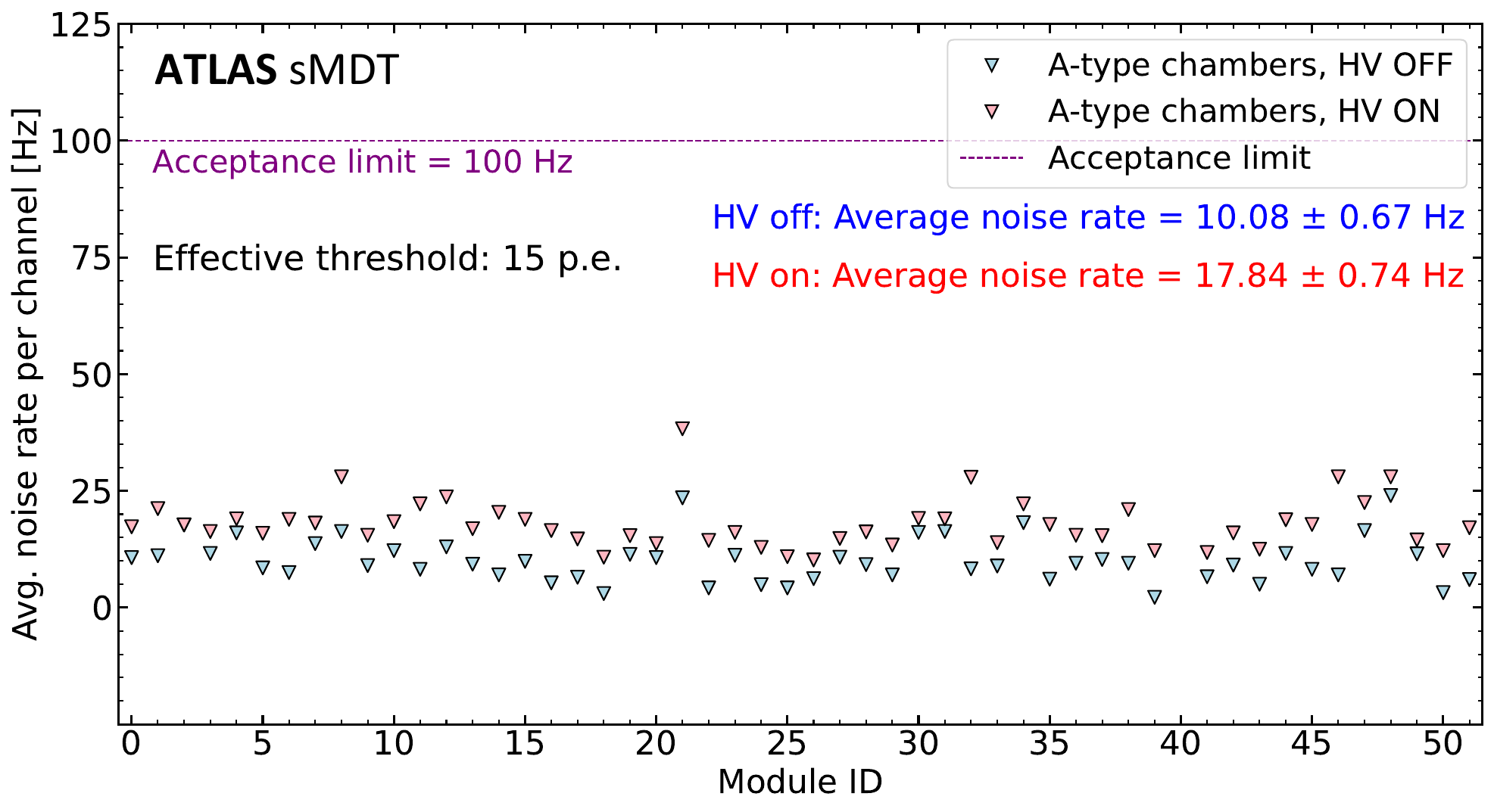}
        \label{fig:noise_Atype_15pe}
        }
    \subfloat[]{       
        \includegraphics[width=0.49\textwidth]{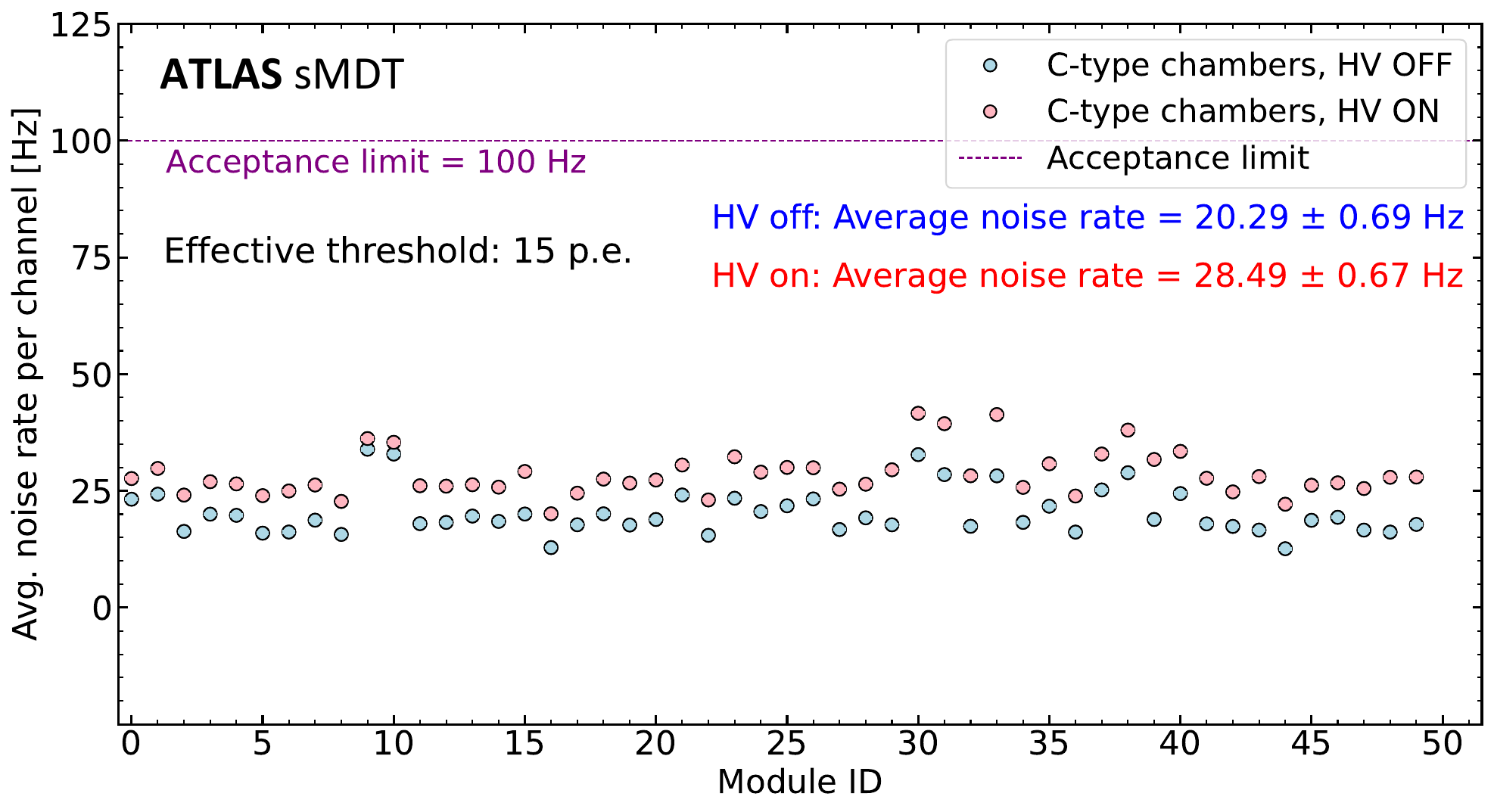}
        \label{fig:noise_Ctype_15pe}
        }\\
     \subfloat[]{   
        \includegraphics[width=0.49\textwidth]{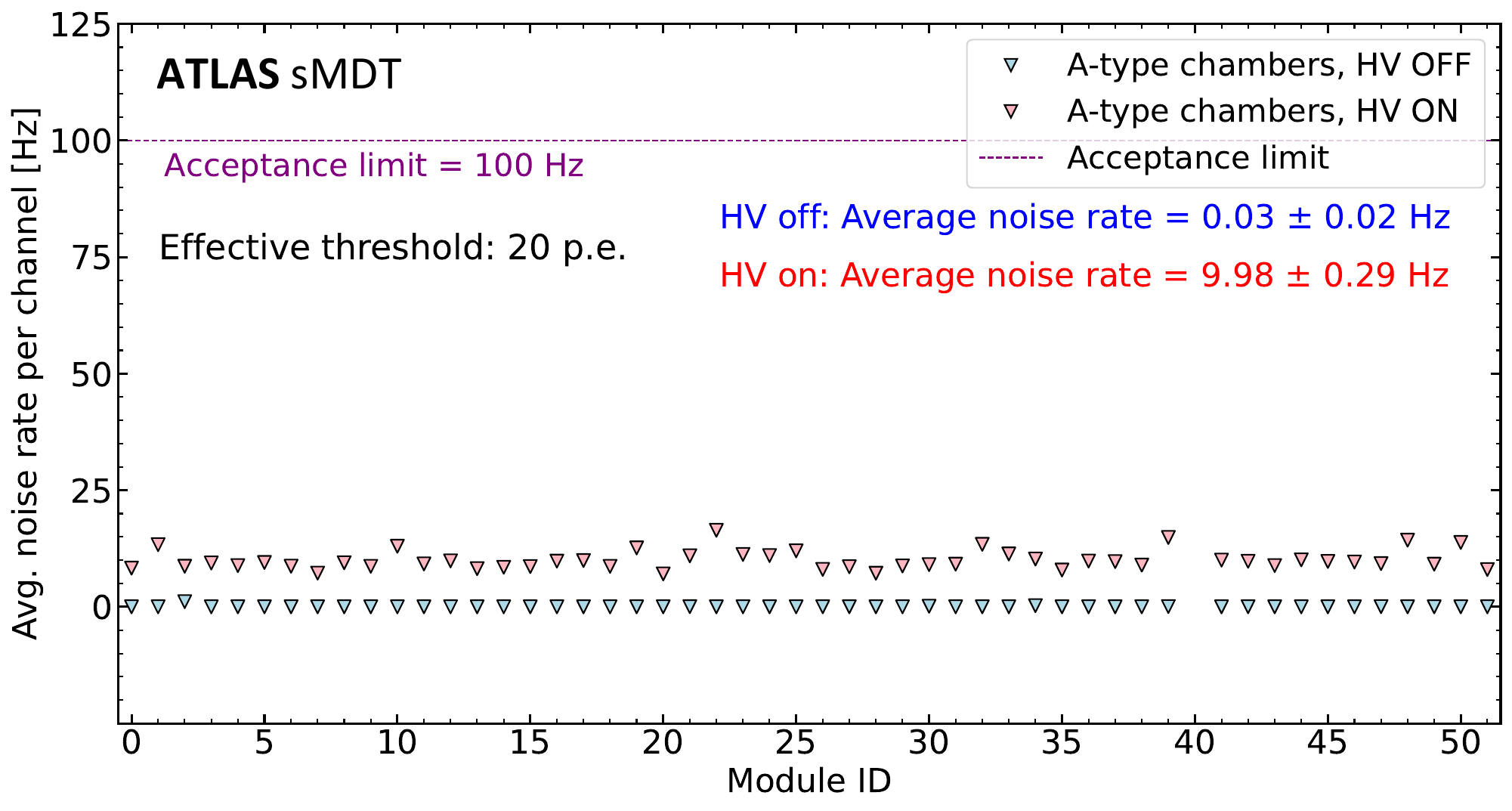}       
        \label{fig:noise_Atype_20pe}
        }
    \subfloat[]{       
        \includegraphics[width=0.49\textwidth]{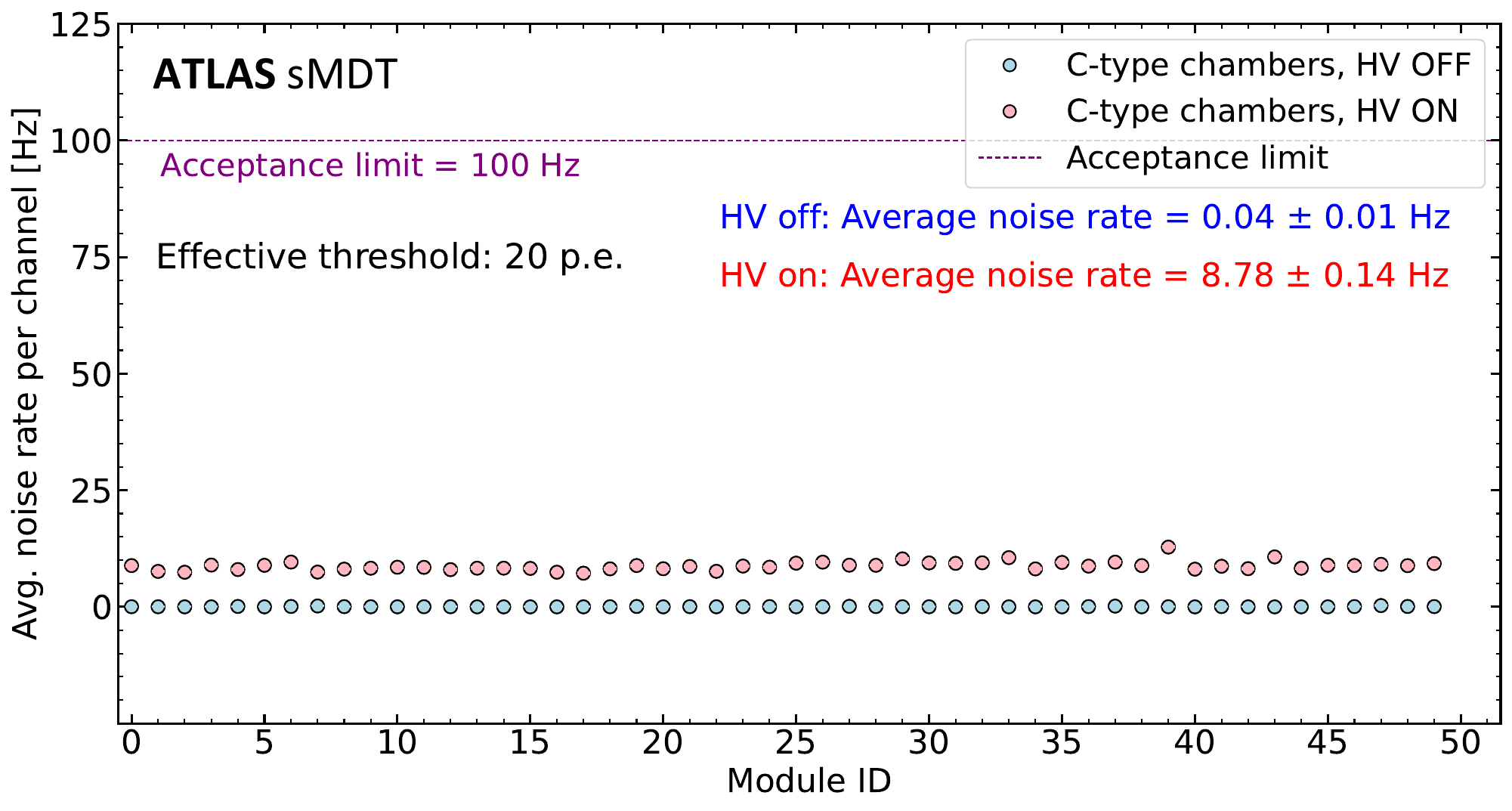}
        \label{fig:noise_Ctype_20pe}
        }       
\caption{Average drift tube noise hit rates of the BIS1--6 sMDT chambers measured with nominal operating voltage turned on and off. (a) A-type and (b) C-type chambers tested at an initially low effective threshold of 15~p.e. for higher sensitivity to channels susceptible to noise.
(c) A-type and (d) C-type chambers tested at the nominal effective threshold of 20~p.e., optimized for safe operation in ATLAS. 
The dashed lines denote the 100~Hz acceptance requirement.}
\label{fig:noise_results}
\end{figure}

At CERN, the chambers were certified again with respect to noise performance (and afterwards for performance with cosmic ray muons) at an effective threshold of 20 primary electrons (p.e.), corresponding to equivalent threshold and hysteresis settings of 114 and 14, respectively, for A-type and 108 and 7 for C-type chambers (see Section~\ref{sub_sec:Readout_elex}).
The noise hit rate per tube was measured with and without the nominal HV of 2730~V applied, using a random trigger, with the chambers operated at the nominal gas mixture of Ar:CO\(_2\) (93:7) at 3~bar absolute pressure. The threshold settings were optimized in a threshold scan (see Figure.~\ref{fig:threshold_scan_opt}), allowing for safe operation in ATLAS well below the exponential turn-on of noise counts. 
The noise hit rate of each tube and the average over all tubes of a chamber at these settings are required to be below \SI{1}{kHz} and \SI{100}{Hz}, respectively.
During all tests, the chambers, the readout electronics, the trigger scintillation counters, and the power supplies shared a common building earth in a star-grounding configuration. 
Initially, the noise hit rates were also evaluated and mitigated where necessary, at a significantly lower effective threshold of 15 p.e., corresponding to threshold and hysteresis settings of 119 and 14 for A-type and 114 and 7 for C-type chambers, for which they fulfilled the requirements as well~(see Figure~\ref{fig:noise_Atype_15pe} and \ref{fig:noise_Ctype_15pe}).

First measurements identified elevated noise hit rates for a small fraction of tubes, which could be attributed to poor ground connections of the tube walls, most common at the chamber edges, producing increased noise even with HV turned off. 
%and intermittent internal sparking caused by dust or water %vapor, observed only with HV on and accompanied by increased %dark current and abnormal ADC spectra.
Within each chamber, grounding screws between every three adjacent tubes contact the tube walls and establish a ground connection from the chambers to the readout electronics~\cite{ATLAS-TDR-026} (see Figure~\ref{stacked_mezzanine}). Some tubes, however, had insufficient mechanical contact 
with the standard grounding screws, 
resulting in increased noise in the adjacent tubes, in particular at the edges of the MLs. Replacing these screws with larger-diameter screws at both readout and HV sides improved the ground connection and reduced the noise substantially. 
%A total of 763 grounding screws were replaced in all chambers at the readout and HV sides.
%In total, 763 screws (410 A-type and 353 C-type) were replaced at %both ends of approximately 390 edge tubes. A few corner tubes with %larger gaps required additional adjustment, adhesive fixation, and %use of even larger-diameter screws.

\if{0}
Tubes contaminated by dust exhibit elevated counting rates and abnormal ADC and TDC spectra, characterized by large-amplitude ADC signals and random TDC hits, consistent with intermittent HV-induced discharges. The affected tubes were conditioned by applying inverted HV of \SI{-3000}{V} until the current stabilized at a few nA, followed by evacuation and refilling with fresh gas. In rare cases, Ar:CO\textsubscript{2} gas at \SI{3}{bar} was flushed directly through the tube to remove contamination.

Water vapor can similarly induce intermittent discharges. At high humidity in the test room, dark currents inside the FCs increase the noise hit rate. They have been reduced to normal levels by flushing dry nitrogen through the FCs and, eventually, by enclosing the chambers in a dehumidified environment at below 40\% relative humidity.
\fi

Removing dust contamination and controlling humidity to a level below 40\% as well as flushing dry nitrogen through the FCs efficiently reduced the chamber leakage currents. The dark currents of each ML were reduced to the range of 20--\SI{50}{nA}, corresponding to an average per tube of less than \SI{0.2}{nA}, well below the requirement of not more than \SI{2}{nA} for individual drift tubes~\cite{ATLAS-TDR-026}.

Following these treatments, all chambers and channels met the noise-rate requirements for HV on and off. The average drift tube noise hit rates of the A- and C-type chambers at low and nominal effective thresholds
are shown in Figure~\ref{fig:noise_results}. 
The results demonstrate the successful mitigation of noisy tubes. The average noise hit rates are well below the required maximum, leaving ample margin for operation in the ATLAS detector. 

\section{Chamber performance evaluation with cosmic rays}
\label{sec:results}
The efficiency and spatial resolution of each drift tube were measured using cosmic-ray data. As shown in Figure~\ref{BB5_teststands}, two chambers are stacked in each teststand underneath a large scintillation counter covering the full chamber width as a trigger detector, enabling simultaneous testing of the pair of chambers. The trigger is provided by the coincidence of signals from photomultiplier tubes attached to both ends of the scintillation counters as described in Section~\ref{sub_sec:Readout_elex}.
The operating conditions used in the cosmic ray tests were identical to those used in the noise hit rate measurements  (Ar:CO\(_2\) 93:7 at 3~bar, HV 2730~V), including the threshold settings for A- and C-type
chambers.
\if{0}
with a main threshold of \SI{-27.00}{mV} and a hysteresis threshold of \SI{8.75}{mV} (corresponding to an effective threshold of \SI{-35.75}{mV}), with at least two million events recorded for each configuration. As noted in Section~\ref{sub_sec:noise_hv}, a different hysteresis setting was used for A-type chamber tests, corresponding to an effective threshold of \SI{-44.50}{mV} instead of \SI{-35.75}{mV}, which also leads to a slightly degraded spatial resolution.
\fi

During chamber construction, 41 non-functional tubes, mostly due to broken wires, were identified and disconnected from gas supply, HV, and readout electronics by cutting the signal pins at both ends and applying HV-insulating glue. Four additional tube failures due to broken wires were found during commissioning at CERN, bringing the total to 45 non-functional tubes (0.09\%).

Both quick online and offline analyses were performed to determine hit efficiency and spatial resolution per tube, allowing early identification of potential deviations in gas composition, pressure, electronics noise, and possible front-end electronics faults.
The ADC spectra of all channels were systematically inspected.
Some of the newly installed mezzanine cards exhibited abnormal signal gains with ADC peaks outside the nominal range. The expected peak of the ADC spectrum for cosmic ray signals is around 150 (see Figure~\ref{fig:spectra_adc}), with some channel-to-channel variations arising from intrinsic ASD chip production differences. Mezzanine cards with channels peaking outside 115--235~ns were replaced, in total 1.6\%, before the final evaluation of the chamber performance.
%Chambers tested before this criterion was established were %subsequently retested, and affected cards replaced. In total, 1.6\% %of the mezzanine cards were replaced across the two teams to ensure %uniform detector response.

The offline data analysis was performed following the procedure described in~\cite{Nelson_2021}. After removing hits with a drift radius deviating from the fitted track by more than five times the hit-radius error, cosmic-ray events with 6 to 12 hits on the fitted track were retained for efficiency and resolution calculations to suppress noise events. The hit efficiency of the drift tubes was evaluated first. It is given by the number of tubes traversed by the fitted tracks with a recorded hit divided by the total number of tubes traversed by the same tracks.

\begin{figure}[!htbp]
    \centering
    \includegraphics[width=0.6\textwidth]{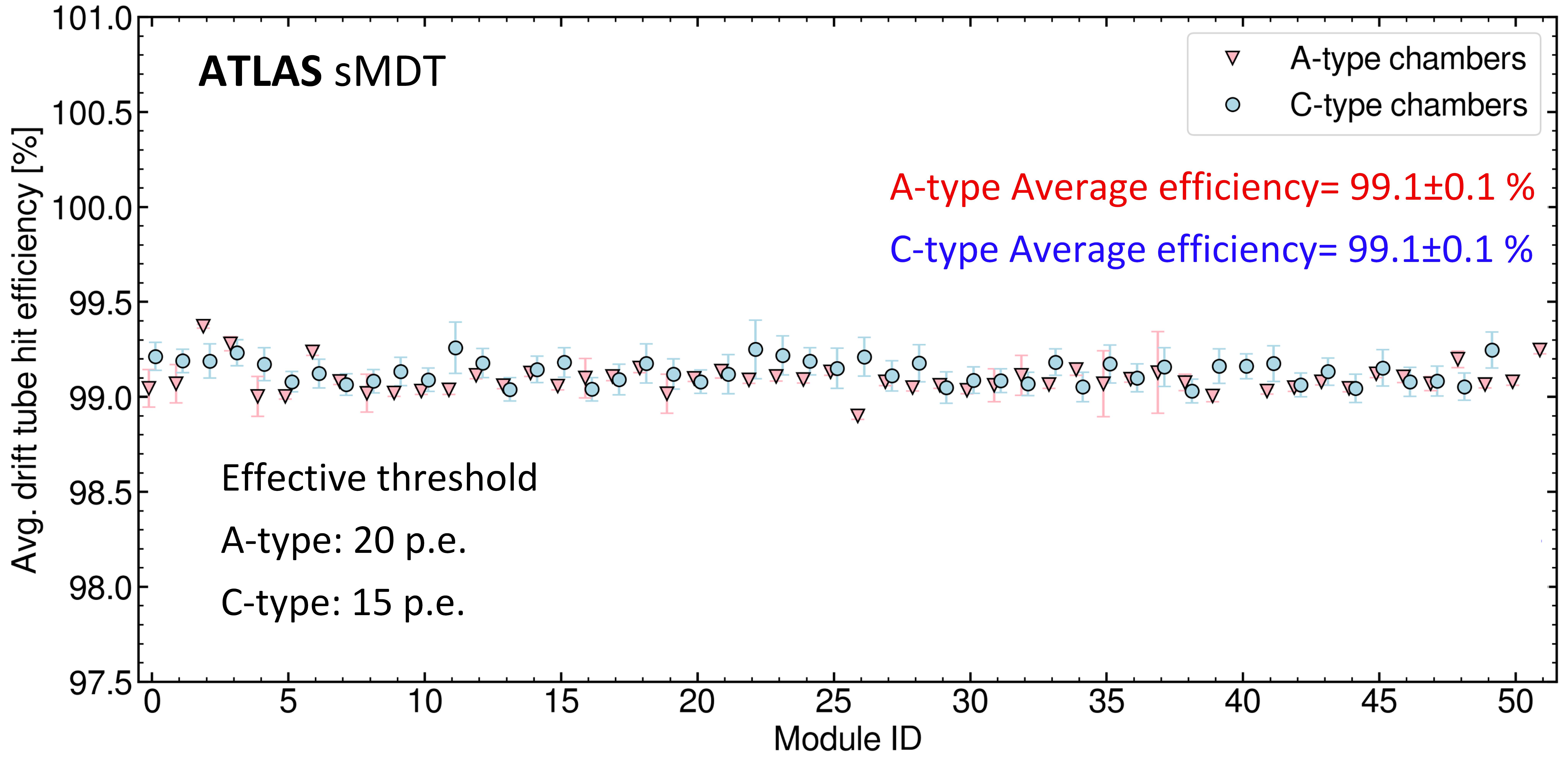}
    \caption{Average drift tube hit efficiency of (a) A-type and (b) C-type BIS1--6 sMDT chambers within the active volume, i.e.\ excluding the tube walls and space between them.}
    \label{fig:eff}
\end{figure}

The average drift-tube hit efficiencies per chamber are summarized in Figure~\ref{fig:eff}. The efficiencies reported here exclude the tube walls and space between them as well as the few dead tubes in the calculation, thereby reflecting the hit efficiencies of active tube volumes. They are stable and uniform over the whole production time, which corresponds to the module number sequence. The average value over all 102 BIS1--6 sMDT chambers is \( (99.1 \pm 0.1)\% \).

For the spatial resolution measurements, time slewing corrections (TSC) are applied first to mitigate the jitter in the threshold crossing time arising from variations in signal amplitude. The correction for each hit is given by~\cite{Nelson_2021}:

\begin{equation}
    \label{eq:TS_corr}
    \Delta t_{\text{slew}} = 35.59 \, e^{-ADC/61.33} \quad [\text{ns}]
\end{equation}
where $ADC$ stands for the measured ADC value of the hit.

\begin{figure}[!hbtp]
    \centering
    \subfloat[]{
        \includegraphics[width=0.5\textwidth,height=4.2cm]{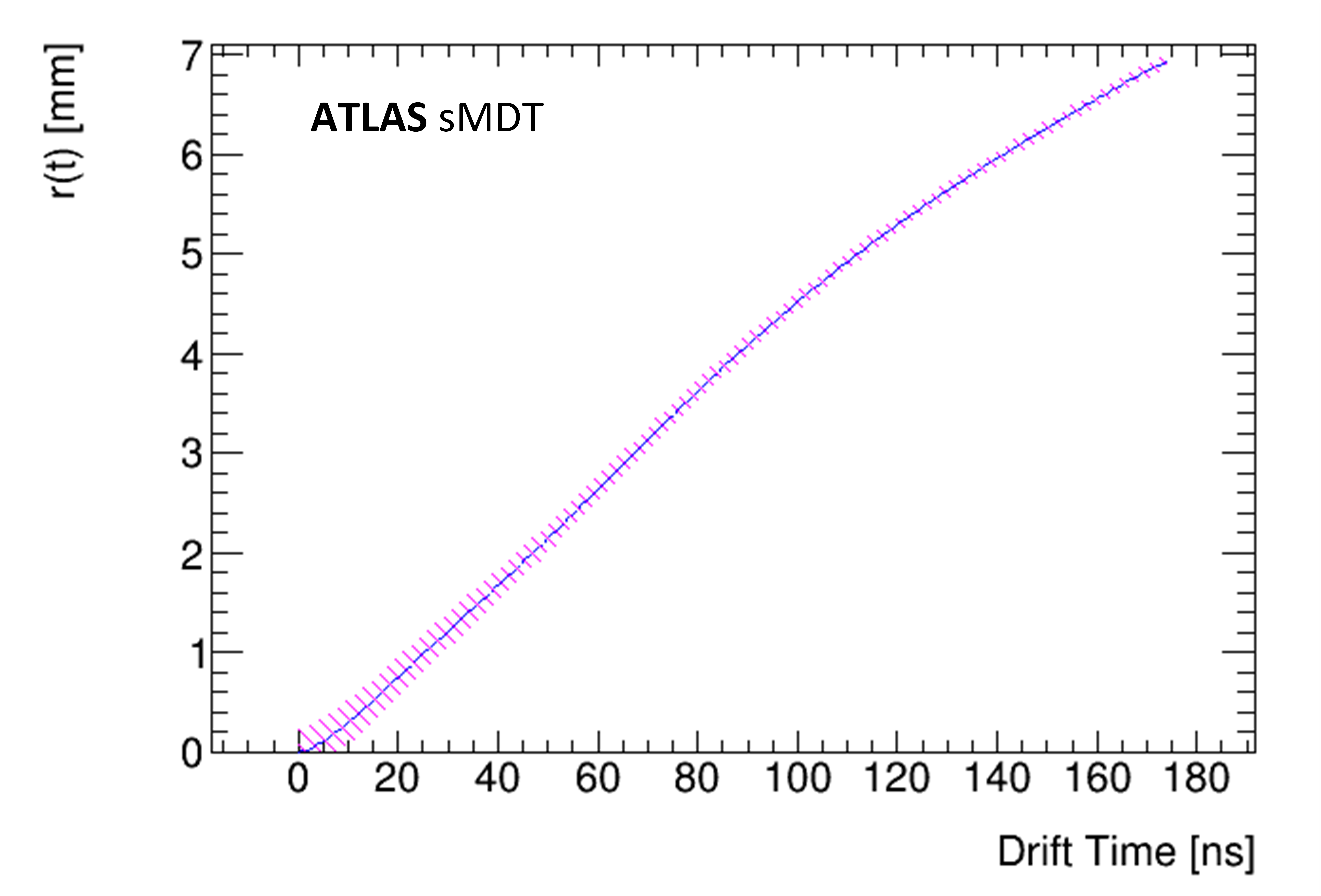}
        \label{fig:rt}
    }
    \raisebox{0.2cm}{
    \subfloat[]{
        \includegraphics[height=4cm]{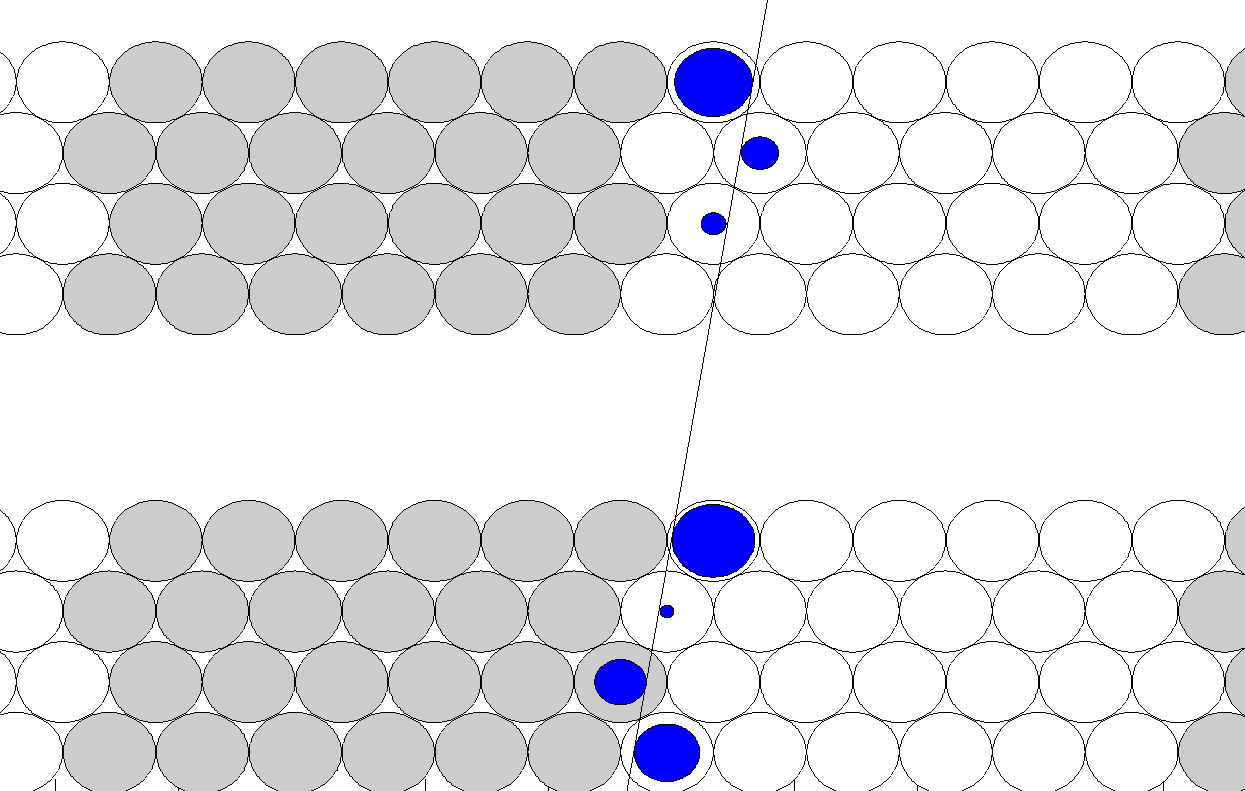}
        \label{fig:event_display}
    }}
    \caption{(a) Example r-t relation of sMDT tubes at the nominal operating conditions, derived from auto-calibration of cosmic-ray data and parametrized by a Chebyshev polynomial.
%    The domain [-1,1] of the Chebyshev polynomial is scaled to [0,175] ns for 
%    illustration. 
(b) Event display showing a reconstructed cosmic-ray muon track in a local sMDT chamber section.
    }
    \label{fig:rt_eventdisplay}
\end{figure}

The rising and falling edges of the corrected drift time spectra are fitted with a Fermi-Dirac function for the leading edge and a parametrization of the falling edge (see Figure~\ref{fig:spectra_tdc}) to determine the minimum and maximum drift times, $t_0$ and $t_{max}$, for each tube. The parameter \(t_0\) at the middle of the rising edge serves as the most precise and consistent time-zero reference accounting for electronic delays and offsets. The scintillator trigger was not used for timing calibration, as its resolution is not sufficient. 
\if{0}
and the drift time of electrons ionized closest to the anode wire. It is derived from a fit to the leading edge of the drift-time spectrum using a Fermi–Dirac step function, defined as follows:

\begin{equation*}
    \label{eq:t0_fit}
    b+\frac{A}{1+e^{-(t-t_0)/T}}
\end{equation*}
where $b, A$, and $T$ are the additional floating parameters in the fit: $b$ is the background noise floor, $A$ is the amplitude, $T$ is the rise time ($t_0$ slope).  In our cosmic-ray setup, the scintillator trigger is not used for timing calibration, as its resolution is insufficient and timing offsets vary between channels. The direct fit to the drift-time spectrum provides a more accurate and self-consistent time-zero calibration. Similarly, the maximum drift time $t_{max}$ is derived by fitting the falling edge.
\fi
An auto-calibration procedure~\cite{Nelson_2021} is performed with cosmic muon tracks to determine the average r-t relation for each ML of a chamber, assuming uniform gas mixture, temperature, and pressure, which is parameterized by a 10\textsuperscript{th}-degree Chebyshev polynomial rescaled to the drift time range. 
%Given that Chebyshev functions are defined on [-1,1], the drift time of each hit is %linearly scaled to this domain with the corresponding $t_0$ and $t_{max}$ of each tube. 
%For each chamber, two r(t) functions are derived, corresponding to the two MLs. 
A typical example is shown in Figure~\ref{fig:rt}. 
%where the time axis is scaled to [0, 175] ns.
 %Using this calibrated r-t relation, the measured drift time of each hit is converted into a drift radius, allowing the reconstruction of muon tracks within the chamber (see Figure~\ref{fig:event_display} for an example cosmic-ray muon track).

With the r-t relations established, the drift radius $r$ of the hit signals is determined from the corrected drift time. A linear fit to the resulting drift circles around the nominal sense wire positions in the drift tube layers reconstructs straight muon trajectories in the plane perpendicular to the tubes (see Figure~\ref{fig:event_display} for an example cosmic-ray muon track).
The distance from the fitted track to the sense wire position defines the predicted drift radius. The difference to the measured drift radius gives the track residual. Both biased and unbiased residuals are determined for each hit. The biased residual is obtained from the fit using all hits associated to the track, whereas the unbiased residual is obtained by refitting the track excluding the hit under consideration. A double-Gaussian function is fitted to each residual distribution to determine the biased and unbiased widths 
$\sigma_b$ and $\sigma_u$, respectively, where the widths of the two Gaussian components are weighted by their amplitudes.
The spatial resolution is then defined as the geometric mean of the two residual distribution widths:
\begin{equation}
    \label{eq:res}
    \sigma=\sqrt{\sigma_b\times\sigma_u}
\end{equation}

\if{0}
\begin{equation*}
    \label{eq:weighted_width}
    \sigma_{b/u}=\frac{A_w\sigma_w+A_n\sigma_n}{A_w+A_n},
\end{equation*}
where $A$ is the amplitude of the Gaussian, $\sigma$ is its standard deviation, and the subscripts $n$ and $w$ are the narrow and wide Gaussians of the double Gaussian distribution. 
The spatial resolution is then defined as the geometric mean 
$\sigma=\sqrt{\sigma_b\times\sigma_u}$ of the two residual distribution  widths.

\begin{equation*}
    \label{eq:res}
    \sigma=\sqrt{\sigma_b\times\sigma_u}.
\end{equation*}
\fi
To account for multiple Coulomb scattering, which is more significant for cosmic rays without a momentum filter than for muons in the ATLAS detector with typically $p_T\geq20$~GeV, the multiple scattering effects in the track residual distributions were deconvoluted using a residual distribution from Monte Carlo simulation of cosmic muon scattering in the teststands~\cite{Nelson_2021}. The widths of the residual distributions after deconvolution are determined as described above, and the intrinsic spatial resolution is then calculated using Eq.~\ref{eq:res}. 

\begin{figure}[htbp]
    \centering
    \includegraphics[width=0.6\textwidth]{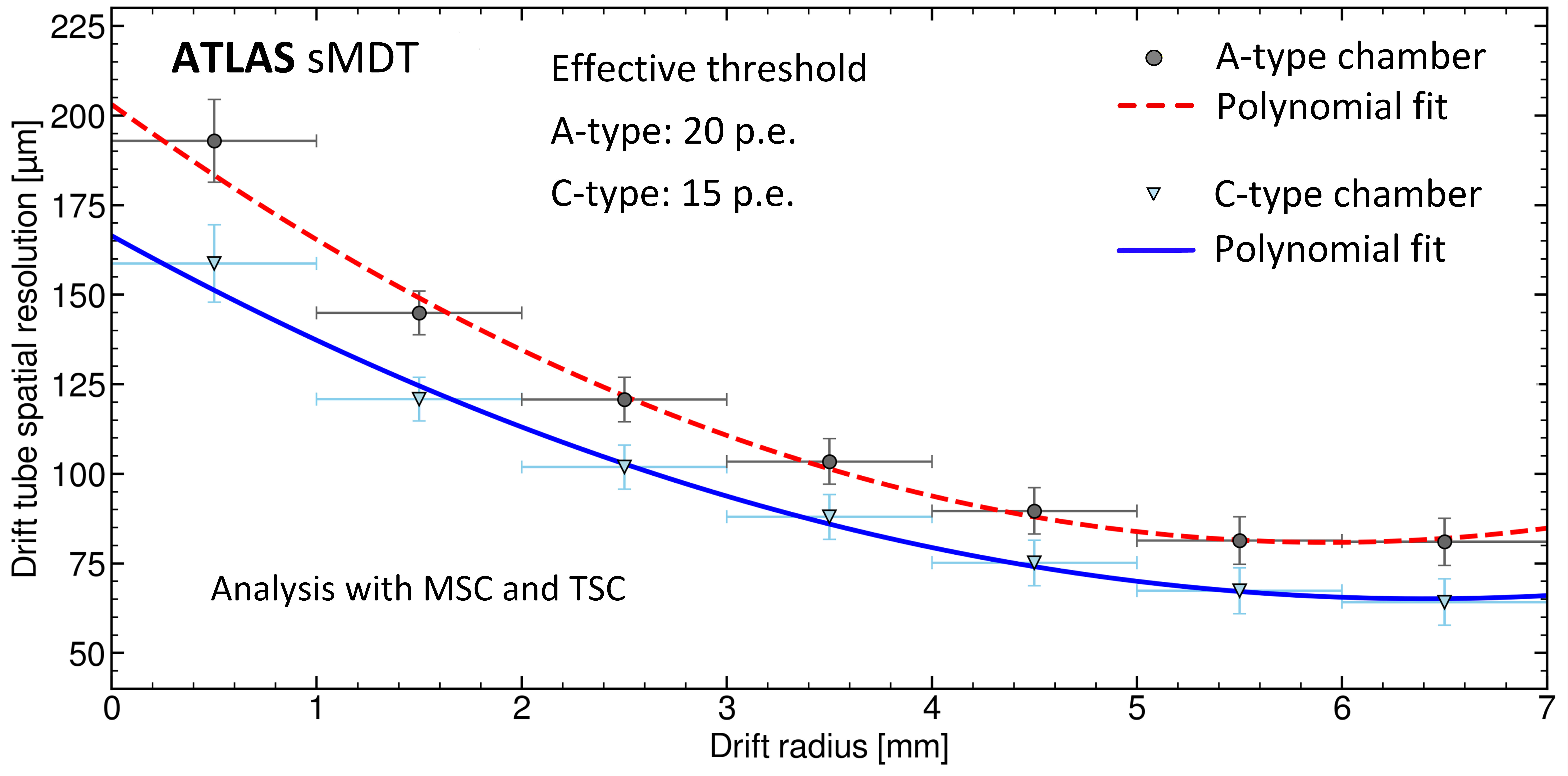}
    \caption{Drift-tube spatial resolution as a function of the drift radius for representative A-type (circles) and C-type (triangles) BIS1--6 sMDT chambers operated at effective thresholds of 20~p.e. and 15~p.e., respectively, including TSC and MSC.}    
    \label{fig:resolution_vs_radius}
\end{figure}

\begin{figure}[htbp]
    \centering
    \subfloat[]{

         \includegraphics[width=0.49\textwidth]{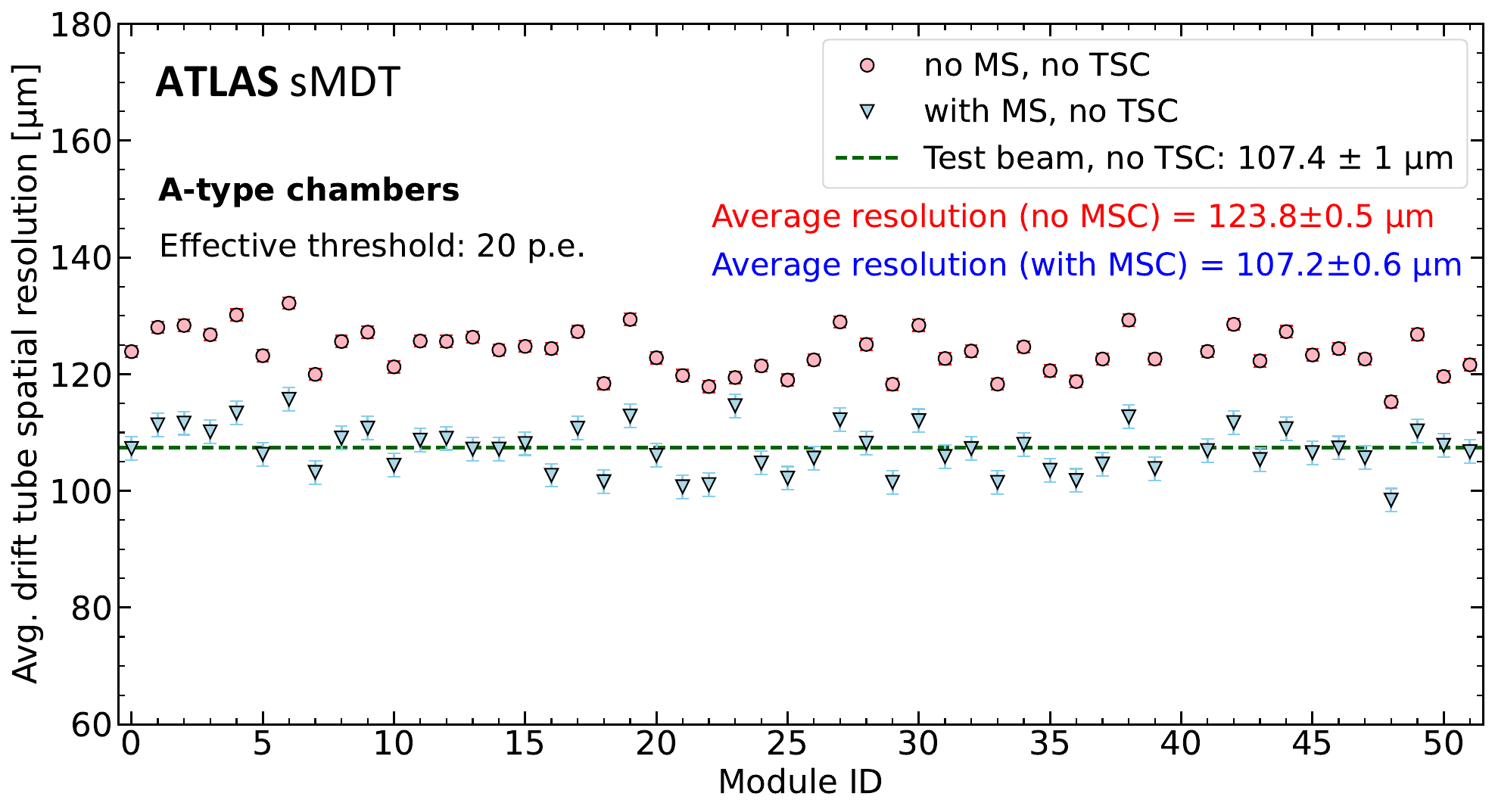}
         \label{fig:resolution_atype_noTSC_20pe}
    }
    \subfloat[]{

        \includegraphics[width=0.49\textwidth]{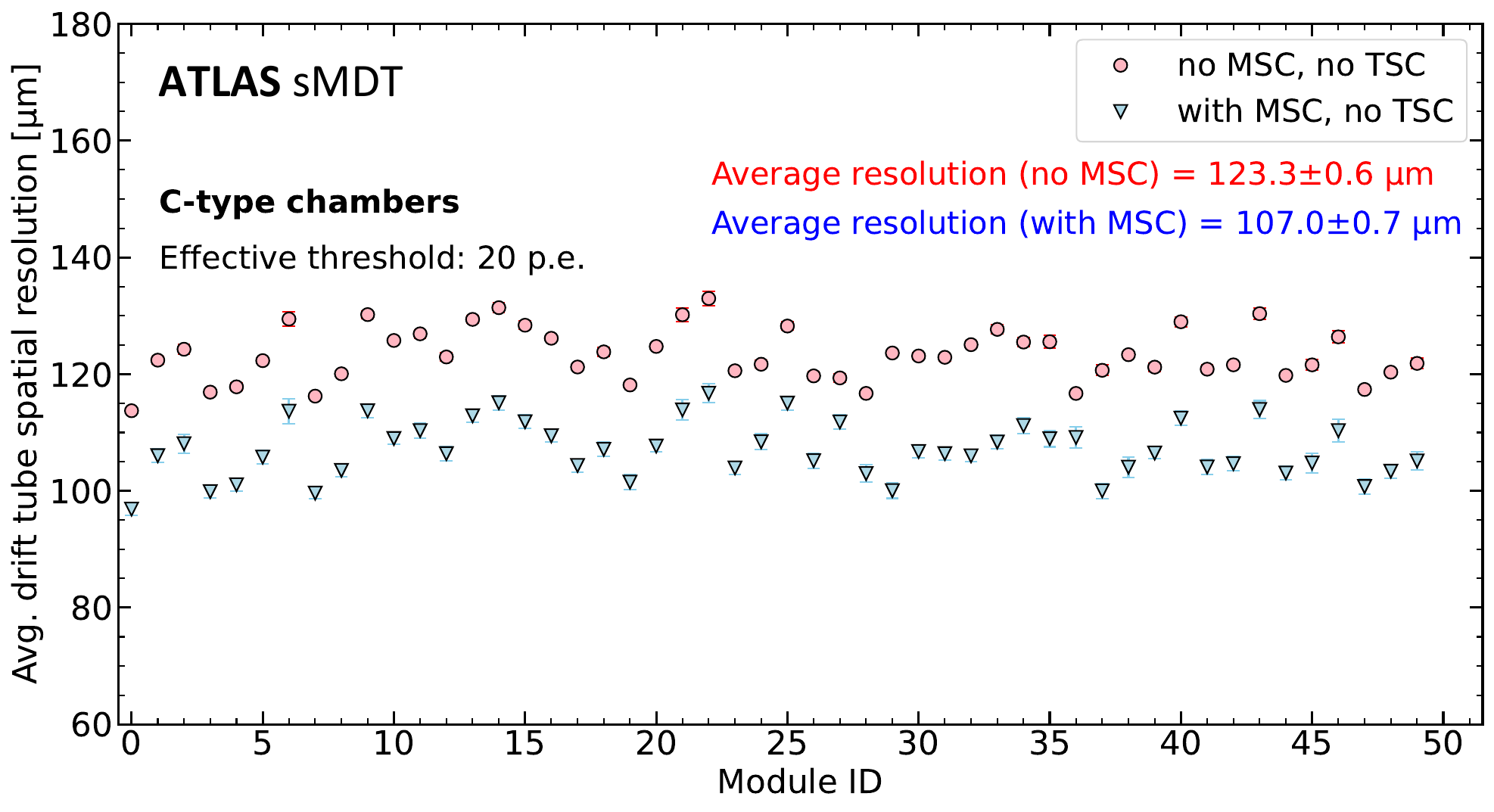}  
        \label{fig:resolution_ctype_noTSC_20pe}
    }\\
    \subfloat[]{

         \includegraphics[width=0.49\textwidth]{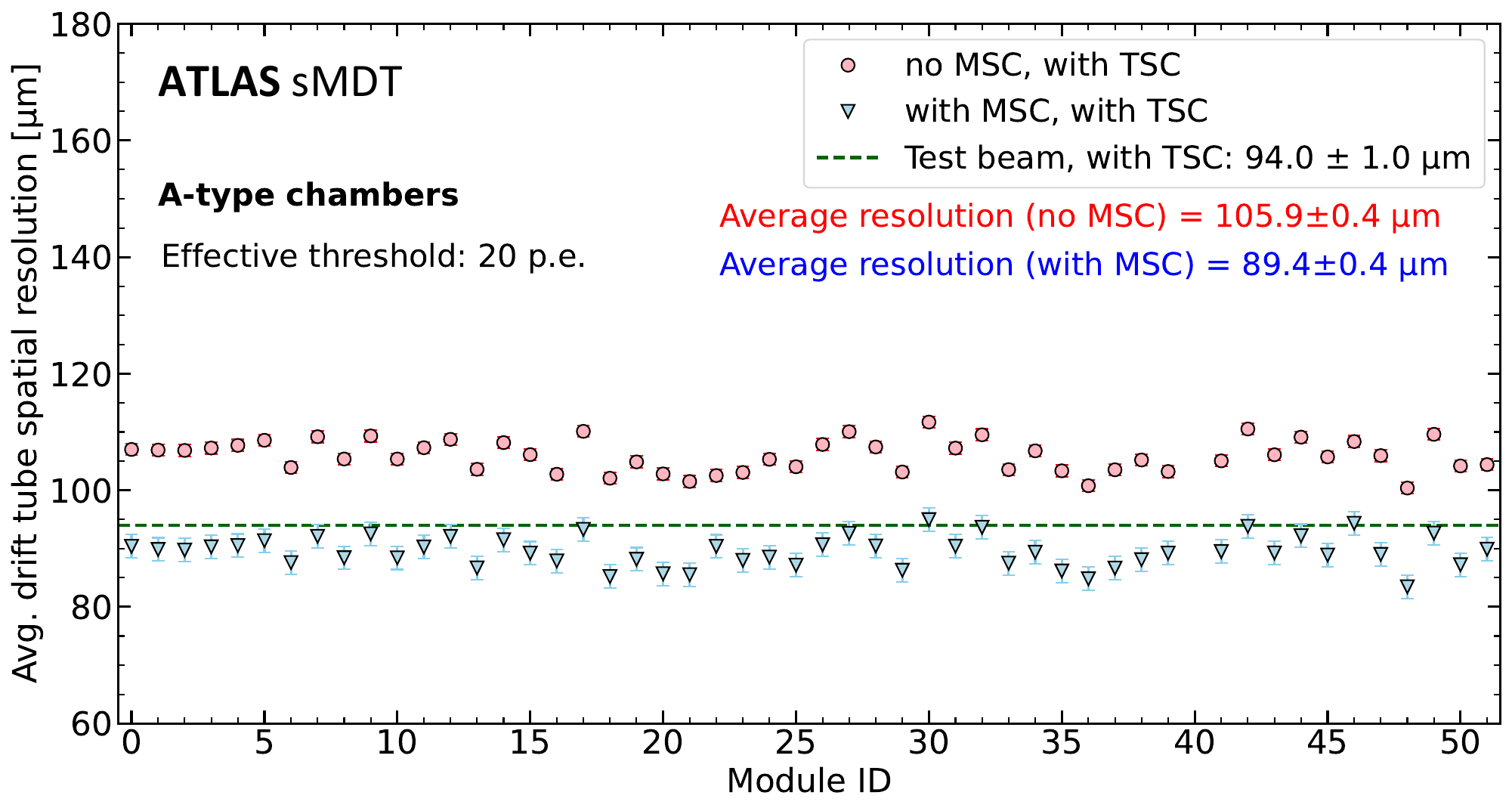}
        \label{fig:resolution_atype_TSC_20pe}
    }
    \subfloat[]{

        \includegraphics[width=0.49\textwidth]{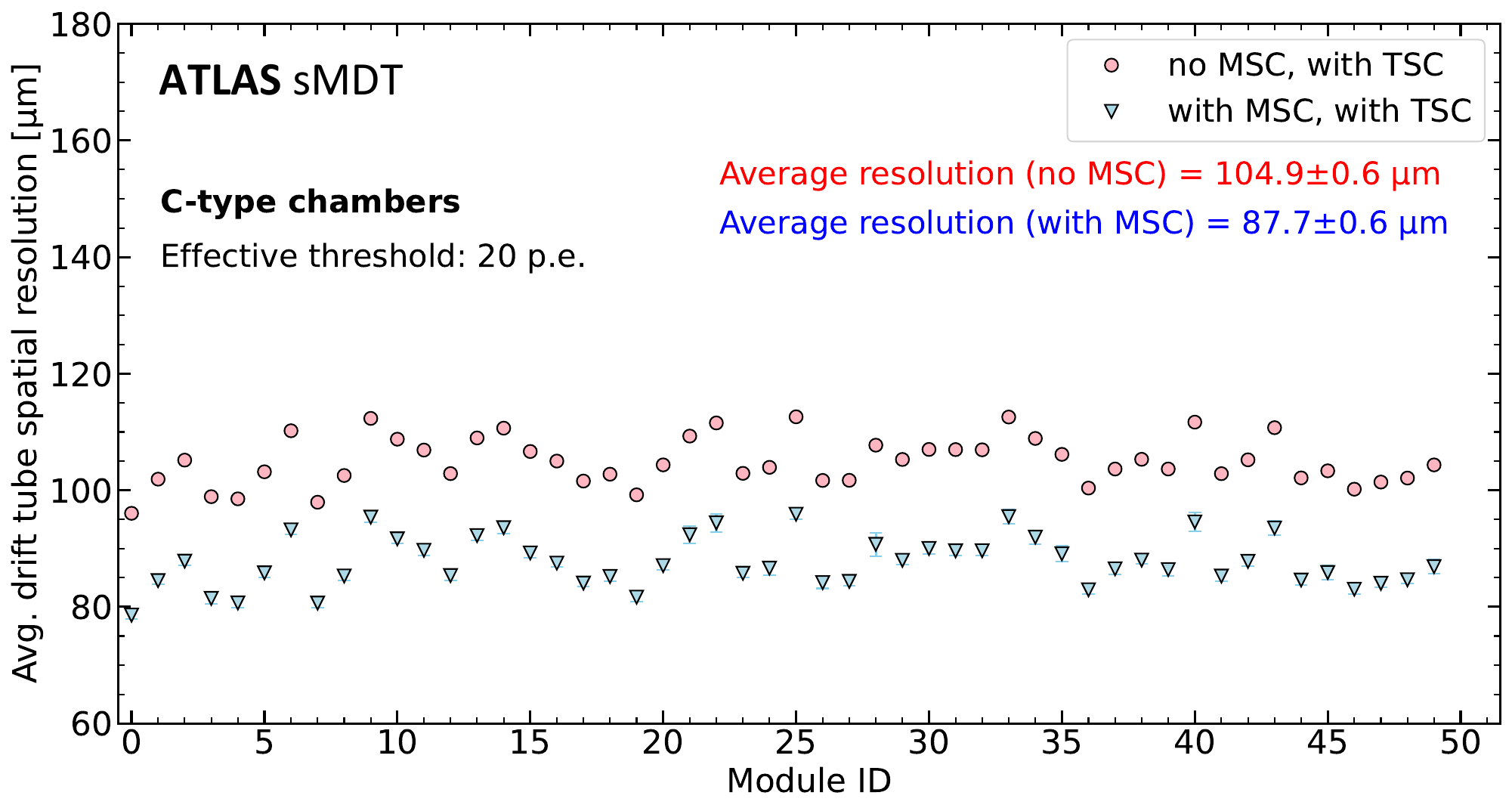}
        \label{fig:resolution_ctype_TSC_20pe}
    }
    \caption{Average drift tube spatial resolutions of the BIS1--6 sMDT chambers for nominal effective threshold 20 p.e., with and without MSC: (a) A-type and (b) C-type chambers without TSC, (c) A-type and (d) C-type chambers with TSC. For comparison, testbeam results obtained at the same effective thresholds with MSC are also shown as dashed lines in (a) and (c).}
    \label{fig:all_res_20pe}
\end{figure}

\begin{figure}[htbp]
    \centering
    \subfloat[]{
        \includegraphics[width=0.49\textwidth]{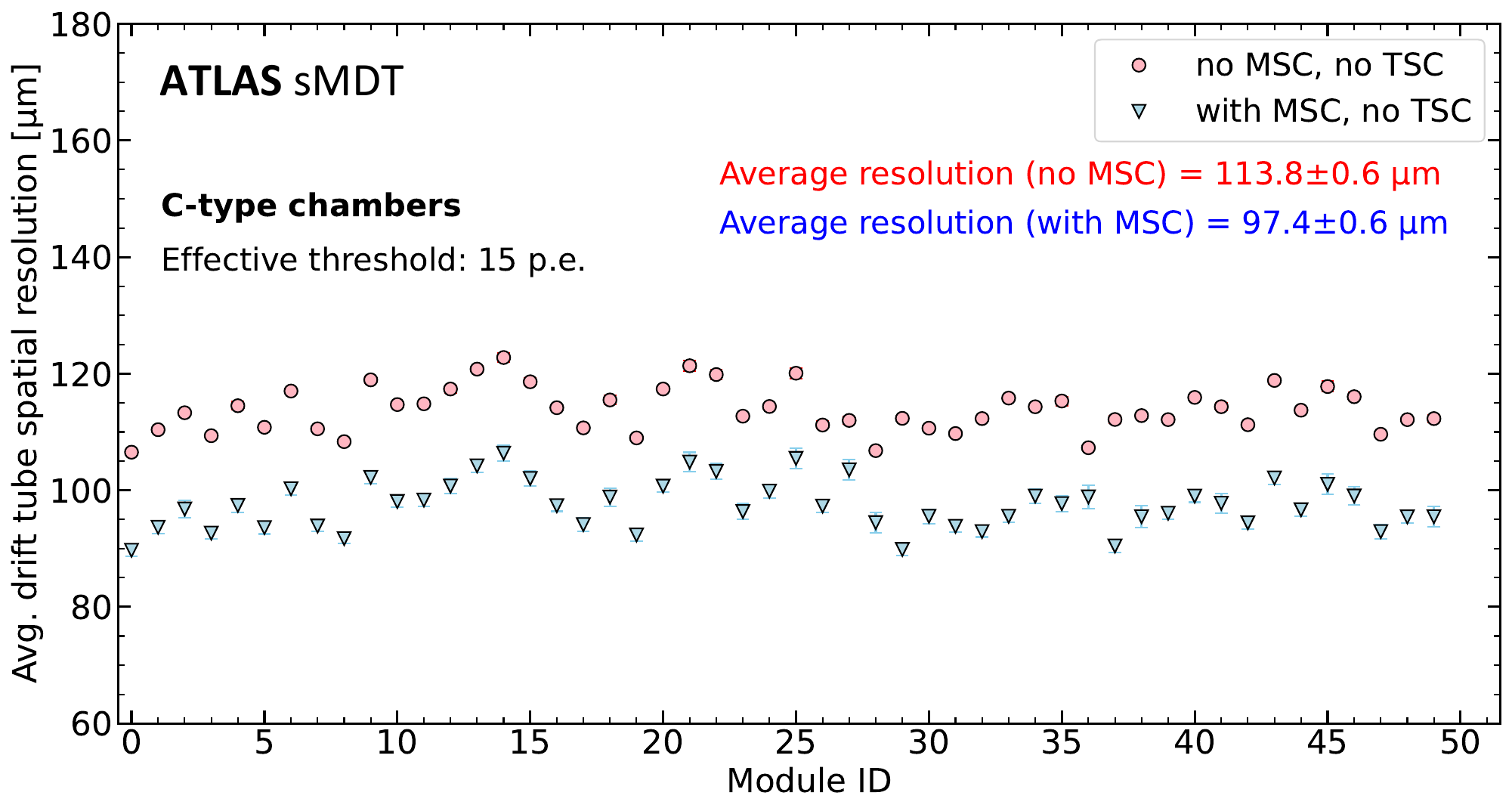}  
        \label{fig:resolution_ctype_noTSC_15pe}
    }
    \subfloat[]{
         \includegraphics[width=0.49\textwidth]{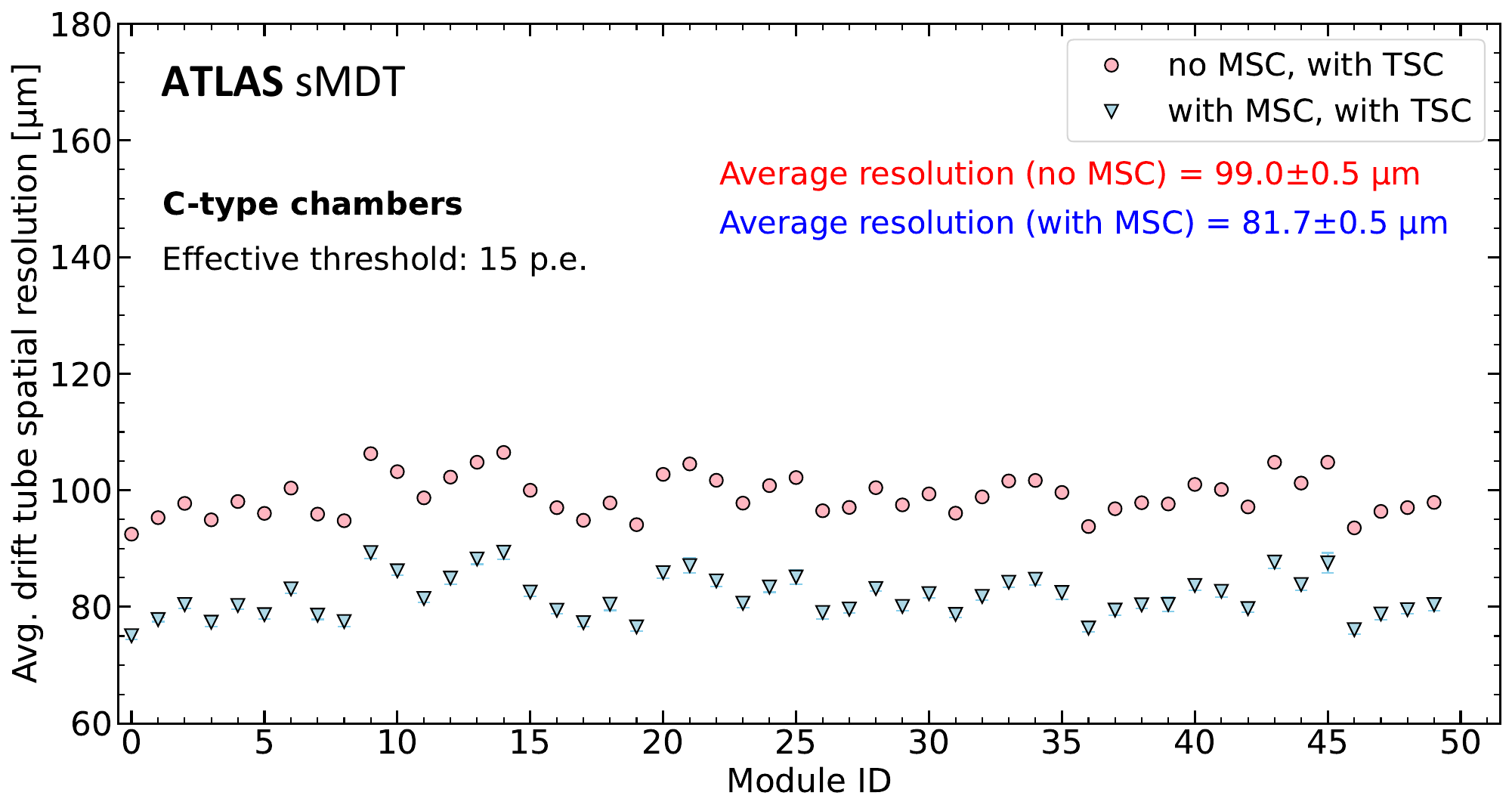}
        \label{fig:resolution_ctype_TSC_15pe}
    }
    \caption{
    Average drift tube spatial resolutions of BIS1--6 sMDT chambers (C-type) for a lower effective threshold of 15 p.e., with and without MSC: (a) without TSC and (b) with TSC.}
    \label{fig:all_res_15pe}
\end{figure}

Figure~\ref{fig:resolution_vs_radius} shows the drift-tube spatial resolution as a function of drift radius for representative BIS1--6 sMDT A-type and C-type BIS1--6 sMDT chambers with both 
TSC and multiple-scattering corrections (MSC) applied. 
The average drift tube spatial resolutions for all A- and C-type chambers, with and without MSC and TSC, are summarized in 
Figures~\ref{fig:all_res_20pe} and \ref{fig:all_res_15pe} for effective thresholds of 20 and 15 p.e., respectively, and in Table~\ref{tab:resolution_summary}.
The resolution results, especially without TSC, demonstrate the stability of the chamber performance and resolution over the production time, which directly corresponds to the module number. They are in very good agreement between A- and C-type chambers for the same threshold settings. The results are also stable over the whole commissioning campaign at CERN. Small variations are due to changes in environmental conditions, especially temperature, not completely compensated by the auto-calibration procedures. 

\begin{table}[!htbp]
\centering
\setlength{\tabcolsep}{4pt}
\renewcommand{\arraystretch}{0.7}
\caption{Summary of drift tube spatial resolutions during sMDT commissioning at CERN, averaged over A- and C-type chambers. The errors are statistical. Testbeam results are also quoted for comparison.}
\label{tab:resolution_summary}
\begin{tabular}{l c c c}
\toprule
Correction condition & \multicolumn{2}{c}{Average spatial resolution ($\mu$m)} & Difference \\
\cmidrule(lr){2-3}
                     & Eff. threshold $20\pm 1$~p.e. & Eff. threshold $15\pm 1$~p.e. & ($\mu$m) \\
\midrule
Without MSC, without TSC & $123.6\pm 0.6$ & $113.8\pm 0.6$ & 9.8 \\
Without MSC, with TSC    & $105.4\pm 0.6$ &  $99.0\pm 0.6$ & 6.4 \\
\midrule
With MSC, without TSC    & $107.1\pm 0.6$ &  $97.4\pm 0.6$ & 9.7 \\
With MSC, with TSC       &  $88.6\pm 0.6$ & $\bf{81.7\pm 0.6}$ & 6.9 \\
\midrule
Testbeam, without TSC    & $107.4\pm 1.0$ &      &   \\ 
Testbeam, with TSC       &  $94.0\pm 1.0$ &      &   \\ 
\bottomrule
\end{tabular}
\end{table}

\begin{table}[!htbp]
\centering
\renewcommand{\arraystretch}{0.8}
\caption{Summary of the sMDT chamber performance parameters other than spatial resolution (see Table~\ref{tab:resolution_summary}) measured in the commissioning campaign at CERN. All values are averages over all A- and C-type chambers. The average drift tube noise rates correspond to HV turned off.}
\label{tab:performance_summary}
\begin{tabular}{l c c}
\toprule
Parameter (chamber average) & Measured value & Requirement \\
\midrule
Chamber gas leak rate [$10^{-3}$~mbar·\text{liter}·s$^{-1}$] & $1.8\pm 0.5$ & $< 9.3$ \\
Tube dark current  & $\sim 0.2$~nA & $< 2$~nA \\
%Noise hit rate per tube & $< 1$~kHz & $< 1$~kHz \\
Tube noise rate (15 p.e.~threshold) & $15$~Hz & $< 100$~Hz \\
Tube noise rate (20 p.e.~threshold) & $0.04$~Hz & $< 100$~Hz \\
Tube hit efficiency (active volume) & $99.1\pm 0.1\%$ & $\ge 99\%$ \\
%Spatial resolution (without TSC) & 99.0~$\mu$m & --- \\
%Spatial resolution (with TSC) & 81.7~$\mu$m & $< 100$~$\mu$m \\
%Total non-functional tubes & 45 (0.09\%) & — \\
\bottomrule
\end{tabular}
\end{table}

Consistency between the resolution results at the production sites and at CERN using the same threshold settings has been shown previously for the A-type chambers based on an earlier commissioning campaign of these chambers in 2023--2024~\cite{KROHA2024169818, BIS1-6_2026}. The improvement of the resolution by TSC amounts to an average shift by $-18.3\pm 1$~\textmu m and $-15.3\pm 1$~\textmu m for effective thresholds of 20 p.e. and 15 p.e., independent of the MSC, which amounts to a uniform shift by $-17$~\textmu m. The TSC is more effective at higher thresholds. The resolutions with MSC obtained for 20 p.e. effective threshold agree well with previous measurements with the same threshold settings in an energetic muon beam~\cite{sMDT_electronics_performance, KROHA2024169818, BIS1-6_2026}, confirming the MSC. The testbeam results, however, correspond to a higher effective threshold of 21~p.e. than used during CERN commissioning, due to the use of ASD chips from pre-production, and have been corrected for this difference. The TSC algorithm has also been improved for the sMDT commissioning compared to the testbeam analysis.


Comparison of the results in Figures~\ref{fig:all_res_20pe} and~\ref{fig:all_res_15pe} demonstrates, with high statistical significance, the dependence of the spatial resolution on the effective threshold. At the same threshold code setting of 114, the effective threshold of 20~p.e. (hysteresis 14) and 15~p.e. (hysteresis 7) yield different resolutions. The drift tube resolution is sensitive to the threshold setting: one threshold code step corresponds to a $1.6~\mu$m change in resolution without TSC and to $1~\mu$m with TSC. For the lower effective threshold of 15 p.e., the average resolution is 99.0~\textmu m after TSC, and improves to 81.7~\textmu m with both TSC and MSC applied, while the noise hit rates remain very low (see Figure~\ref{fig:noise_Atype_15pe} and \ref{fig:noise_Ctype_15pe}). This is about 20~\textmu m better than with the legacy ASD chip at comparable noise hit rates.

\section{Summary}
\label{sec:conclusions}
The surface commissioning of 102 sMDT chambers, 52 A-type and 50 C-type, for installation in the two hemispheres of the ATLAS muon spectrometer in the innermost barrel layer during the upgrade for HL-LHC has been completed at CERN during 2025, following the arrival of the final front-end electronics cards in November 2024. A comprehensive test program has been performed, including gas leak rate, HV dark current, and noise hit rate measurements, as well as tests of the in-plane alignment system and evaluation of hit efficiency and spatial resolution with cosmic rays. %Occasional faults of individual channels after the shipment %to CERN, such as gas leaks, faulty electronics boards, %noisy or sparking tubes, and blocked alignment lines, have %all been successfully resolved during the campaign.

All chambers meet or exceed the stringent requirements. A summary of the measured performance parameters is given in Table~\ref{tab:performance_summary}. In particular, dark currents of the chambers are at a level of 0.2\,nA per tube, an order of magnitude below the 2\,nA limit, gas leak rates per chamber are typically a factor of five below the requirement of \(9.3 \times 10^{-3}~\text{mbar·liter·s}^{-1}\), and the average drift tube noise hit rate per chamber for a discriminator threshold corresponding to 15 p.e. is at a level of 20~Hz, which is a factor of five below the required 100\,Hz limit, while no individual channel exceeds 1\,kHz.  The average drift tube hit efficiency in the active volume is 99.1\%. The average drift tube spatial resolution, corrected for multiple scattering effects, was determined to be $97\pm 0.6$~\textmu without and 
$82\pm 0,6$~\textmu m with time slewing 
corrections at the low noise rates at 15 p.e. threshold. The results are consistent for all the 102 chambers tested. Only 0.09\% of 49152 drift tubes are non-functional and have been disconnected from gas and electronics. The sMDT chambers satisfy all performance criteria to provide precision muon tracking at HL-LHC. The chambers have been tested several times over the past six years since the construction time and showed remarkably stable performance. The next steps are the integration and commissioning with the RPC triplets, the installation of the global alignment and the magnetic field sensors, and finally, the installation and commissioning in the ATLAS cavern.

\acknowledgments
We thank CERN for providing the facilities and infrastructure essential to the sMDT surface commissioning work presented in this study. We are grateful to Spyridon Kompogiannis (Aristotle University of Thessaloniki) and Markus Lippert (MPI Munich) for their assistance with mechanical work at CERN. We also thank Robert Richter (MPI Munich) for explaining the detailed properties of the new ASD ASIC and critically reading the manuscript, Gia Khoriauli (University of Würzburg) for the laboratory test of the new mezzanine cards prior to their installation on the chambers, and Karol Poplawski (NIKHEF) for his support with the RASNIK software upgrade at CERN. 

We thank Claire Lundy, Francie Wharton, and Andrew Kiesling, supported by the NSF--CERN program "Research Experiences for Undergraduates", for their contributions during the summer of 2025. Finally, we thank Professors Jianming Qian and Thomas Schwarz (UM), Reinhard Schwienhorst (Michigan State University), and Hugo Beauchemin (Tufts University) for their continued support of the postdoctoral researchers and graduate students who contributed to this work.

This work is supported by the U.S. National Science Foundation under Award No.~PHY-1948993.

% Bibliography
\bibliographystyle{JHEP}
\bibliography{biblio.bib}

\end{document}